\makeatletter
\@ifundefined{@parse@version@dash}{%
	\def\@parse@version#1{\@parse@version@0#1}
	\def\@parse@version@#1/#2/#3#4#5\@nil{%
		\@parse@version@dash#1-#2-#3#4\@nil}
	\def\@parse@version@dash#1-#2-#3#4#5\@nil{%
		\if\relax#2\relax\else#1\fi#2#3#4 }
}{}
\makeatother

\makeatletter
\DeclareFontFamily{U}{tipa}{}
\DeclareFontShape{U}{tipa}{m}{n}{<->tipa10}{}
\newcommand{\arc@char}{{\usefont{U}{tipa}{m}{n}\symbol{62}}}%

\def\d{{\textrm d}}

\def\n{{\mathbf{n}}}

\def\half{{\textstyle \frac{1}{2}}}

\newcommand{\arc}[1]{\mathpalette\arc@arc{#1}}

\newcommand{\arc@arc}[2]{%
	\sbox0{$\m@th#1#2$}%
	\vbox{
		\hbox{\resizebox{\wd0}{\height}{\arc@char}}
		\nointerlineskip
		\box0
	}%
}
\makeatother

\documentclass[%
reprint,
superscriptaddress,
amsmath,amssymb,
aps,
prl,
showkeys
]{revtex4-2}
\usepackage{verbatim}


\usepackage{color}
\usepackage{graphicx}
\usepackage{dcolumn}
\usepackage{bm}
\usepackage{hyperref}
\newcommand{\beq}{\begin{equation}}
\newcommand{\eeq}{\end{equation}}
\newcommand{\beqs}{\begin{eqnarray}}
\newcommand{\eeqs}{\end{eqnarray}}
\newcommand{\T}{{\rm T}}

\usepackage{amsmath}

\newcommand{\bfa}{{\bf a}}
\newcommand{\bfb}{{\bf b}}

\newcommand{\bfe}{{\bf e}}

\newcommand{\bfk}{{\bf k}}
\newcommand{\bfm}{{\bf m}}
\newcommand{\bfn}{{\bf n}}

\newcommand{\bfr}{{\bf r}}

\newcommand{\bfu}{{\bf u}}
\newcommand{\bfv}{{\bf v}}

\newcommand{\bfx}{{\bf x}}
\newcommand{\bfy}{{\bf y}}

\newcommand{\bfz}{{\bf z}}

\newcommand{\bfF}{{\bf F}}

\newcommand{\bfI}{{\bf I}}

\newcommand{\bfM}{{\bf M}}

\newcommand{\bfP}{{\bf P}}

\newcommand{\bfR}{{\bf R}}
\newcommand{\bfS}{{\bf S}}

\newcommand{\bflambda}{{\boldsymbol{\Lambda}}}
\def\d{{\textrm d} }

\begin{document}

\title{Surface instability in a nematic elastomer}

\author{Morgan Barnes}
\thanks{These two authors contributed equally.}
\author{Fan Feng}
\thanks{These two authors contributed equally.}
\affiliation{Department of Engineering, University of Cambridge, Trumpington St., Cambridge CB2 1PZ, United Kingdom}
\author{John S. Biggins}
\email{jsb56@cam.ac.uk}
\affiliation{Department of Engineering, University of Cambridge, Trumpington St., Cambridge CB2 1PZ, United Kingdom}

\date{\today}

\begin{abstract}
Liquid crystal elastomers (LCEs) are soft phase-changing solids that exhibit large reversible contractions upon heating, Goldstone-like soft modes and resultant microstructural instabilities. We heat a planar LCE slab to isotropic, clamp the lower surface then cool back to nematic. Clamping prevents macroscopic elongation, producing compression and microstructure. We see that the free surface destabilizes, adopting topography with amplitude and wavelength similar to thickness. To understand the instability, we numerically compute the microstructural relaxation of a ``non-ideal" LCE energy. Linear stability reveals a Biot-like scale-free instability, but with oblique wavevector. However, simulation and experiment show that, unlike classic elastic creasing, instability culminates in a cross-hatch without cusps or hysteresis, and is constructed entirely from low-stress soft modes.
\end{abstract}


\maketitle
Liquid crystal elastomers (LCEs) are actuating solids that recall the dramatic shape transformations of biological tissues.  Microscopically, LCEs are networks of LC polymers \cite{warner2007liquid}. Actuation occurs via the isotropic-nematic phase transition, which biases conformations along the director, generating large elongations on cooling, and  muscular contraction on heating \cite{kupfer1991nematic, de1997artificial}. Correspondingly, LCEs fabricated with spatial director profiles can actuate into complex surfaces \cite{white2015programmable} such as cones \cite{modes2011gaussian, guin2018layered} or faces \cite{aharoni2018universal,barnes2019direct}. The symmetry breaking character of the isotropic-nematic transition also endows LCEs with Goldstone-like ``soft-modes" in which deformations induce director rotation at almost zero energy/stress \cite{golubovic1989nonlinear,warner1994soft}. Such modes enable actuation by modest electric fields \cite{urayama2006deformation, corbett2009deformation}  and generate martensitic \cite{bhattacharya2003microstructure}  microstrucural instabilities \cite{verwey1996elastic, finkelmann1997critical, desimone2002macroscopic}. However, despite progress in idealized cases \cite{desimone2002macroscopic, conti2002semisoft, biggins2009textured},  we still lack a realistic coarse-grained constitutive model that accounts for microstructure, preventing analysis of macroscopic LCEs under load.  


Mechanical instabilities like buckling provide an attractive additional route to complex morphing without  pre-patterning. For example, the  Biot creasing instability sculpts cusped folds at the free surface of soft solids under compression \cite{biot1965mechanics,trujillo2008creasing, hohlfeld2011unfolding, tallinen2013surface}, and underpins morphogenesis of   villi \cite{shyer2013villification} and sulci/gyri \cite{tallinen2014gyrification}. Here, we combine experiment, theory and computation to tackle the analogous compressive surface instability in LCEs, which is profoundly enriched  by soft modes and microstructure.

\begin{figure}[t]
	\includegraphics[width=\columnwidth]{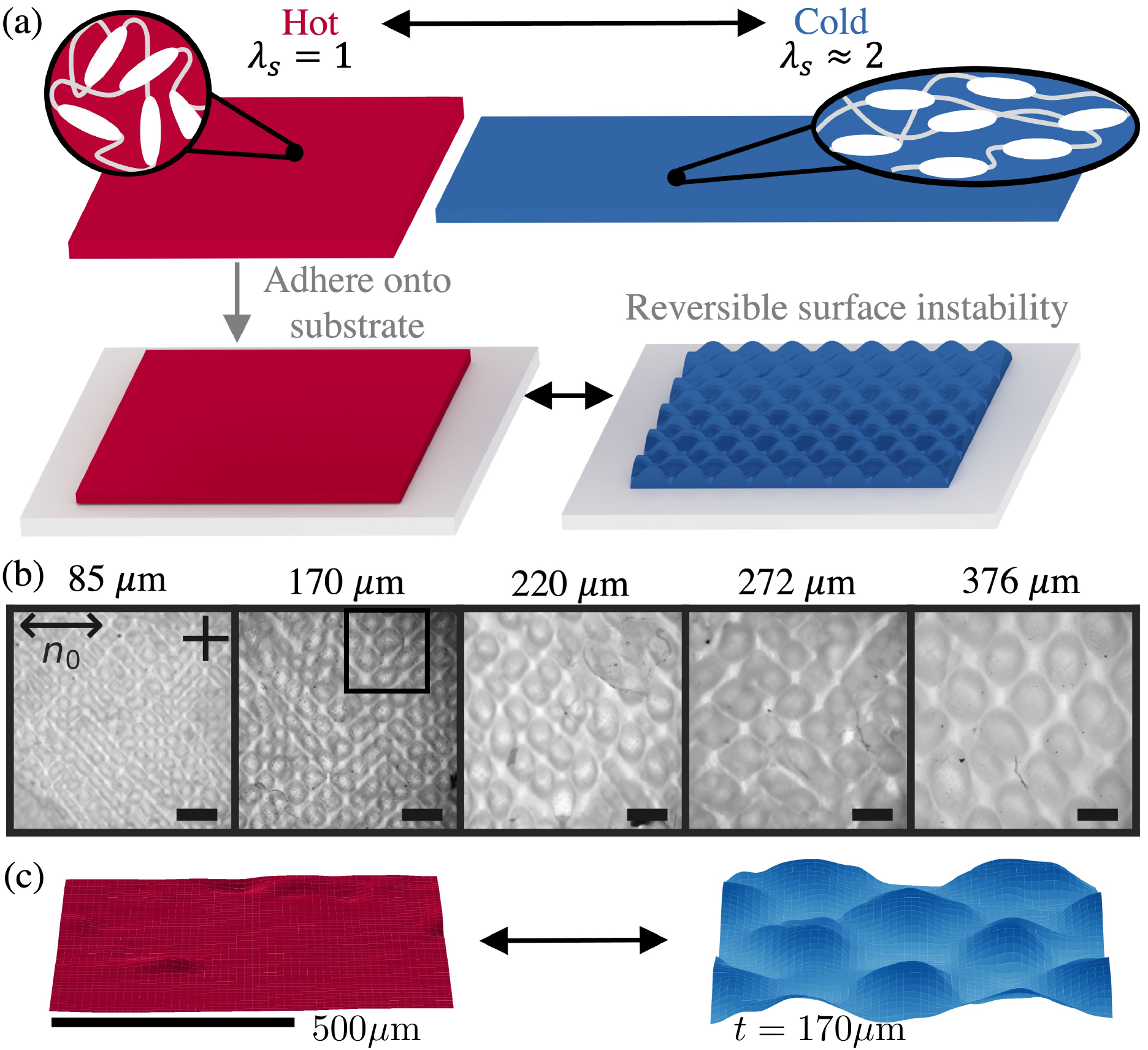}
	\caption{(a) Schematic of LCE actuation between hot (isotopic) and cold (nematic), and resulting surface instability when adhered to a foundation whilst hot. (b) Micrographs (Polarized Optical Microscopy (POM)) of  surface instability for varying LCE thicknesses. (Scale bars 500$\mu$m.) (c)  Topography changes (boxed region)  from optical profliometry (SI Sec.\ III), true aspect ratio.}
	\label{fig:fig1}
\end{figure} 
Experimentally, we fabricate a  monodomain LCE (SI  \cite{supp} Sec.\ I) via two-step cross-linking with mechanical programming \cite{kupfer1991nematic}, using commercially available acrylate-functionalized mesogens and thiol functionalized chain-extenders/crosslinkers \cite{barnes2019direct, yakacki2015tailorable}.  The resulting film is clear but birefringent at room temperature, reflecting its planar nematic alignment, and  elongates reversibly by a factor of $\lambda_s\approx2$ on cooling from isotropic to nematic (SI Fig. S2).  Inspired by observations of creasing in gel slabs that are swollen on a clamped foundation \cite{trujillo2008creasing, tallinen2013surface}, we adhere the LCE to a glass slide whilst isotropic then cool to nematic (Fig.\ \ref{fig:fig1}(a)). The slide prevents elongation, placing the LCE in compression,  causing turbidity in the LCE indicative of optical-scale microstructure. Moreover, the surface destabilizes, forming a high amplitude cross-hatch pattern of topography with wide bumps separated by narrow valleys (Fig.\ \ref{fig:fig1}(b,c), and SI \cite{supp} Sec.III, M1, M2). Topography is reversible on heating (SI \cite{supp} M3), and experiments varying LCE thicknesses show amplitude and wavelength are proportional to thickness. 

To explain the instability,  we first model the LCE as a simple rubber slab undergoing a spontaneous elongation whilst clamped below, leading to a displacement field $\mathbf{u}$ (from isotropic) and  deformation gradient $\boldsymbol{\Lambda}=\bfI+\nabla \mathbf{u}$. A standard neo-Hookean rubber with shear modulus $\mu$ stores elastic energy density $W_{NH}(\boldsymbol{\Lambda})=\half \mu \mathrm{Tr}(\boldsymbol{\Lambda} \boldsymbol{\Lambda}^\T)$, and is strictly incompressible $\mathrm{Det}{(\boldsymbol{\Lambda})}=1$. Accordingly, the LCE energy is $W_{NH}(\boldsymbol{\Lambda}\boldsymbol{\Lambda}_s^{-1})$, where $\boldsymbol{\Lambda}_s(\mathbf{n}) = \lambda_s \n \otimes \n +\lambda_s^{-1/2} (\bfI-\n\otimes \n)$ is the spontaneous elongation. Minimizing this energy in an infinite planar slab with a clamped foundation is known to generate a standard Biot-type surface instability, with cusped  furrows (creases) appearing sub-critically at the linear threshold of $\lambda_s\approx 2.27$ \cite{biot1965mechanics,hong2009formation}, and being global minimizers beyond the non-linear nucleation threshold of $\lambda_s\approx 1.77$ \cite{hong2009formation, hohlfeld2011unfolding} (SI Sec.\ V \cite{supp}). We confirm these expectations with 3D finite elements, using a bespoke C-code from \cite{tallinen2013surface} that divides the slab into constant strain tetrahedra, and moves nodes via damped Newtonian dynamics. These calculations exhibit regular crease-lines perpendicular to the director, agreeing with previous elastomer experiments \cite{cai2012creasing} and numerics \cite{tallinen2013surface}, but disagreeing with the smooth cross-hatch in LCEs. 

\begin{figure}[t]
	\includegraphics[width=\columnwidth]{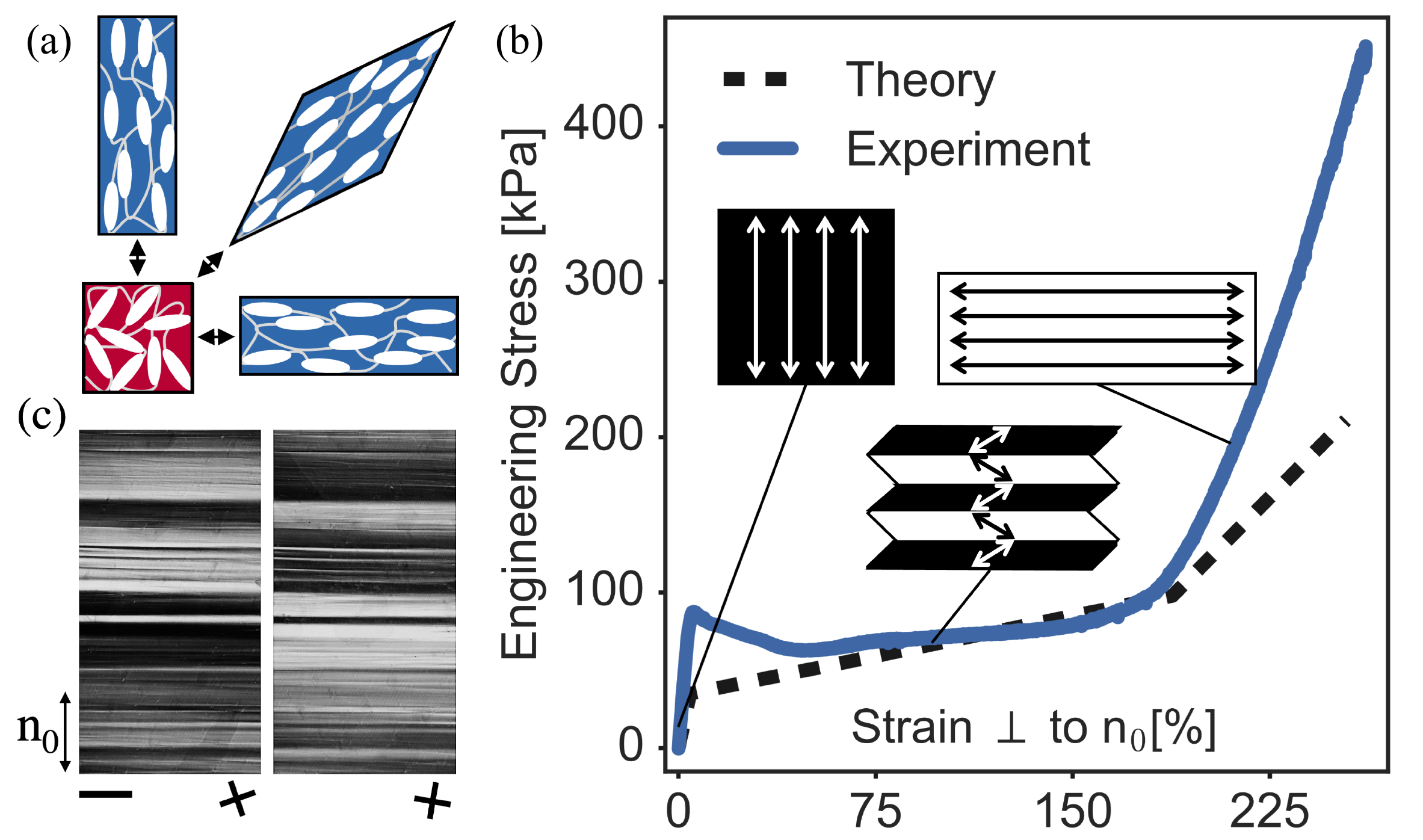}
	\caption{(a) Soft modes in LCEs. (b) Experimental and theoretical \cite{warner2007liquid} stress-strain curves. Theory parameters: $\lambda_s=2, \alpha=0.05, \mu = 600 ~\text{kPa}$ .  (c) Micrographs of a stripe microstructure rotated under cross-polarizers. Scale bar 500$\mu$m. }
	\label{fig:fig2}
\end{figure}

The key additional consideration is that $\n$ can rotate within the LCE, enriching the energy with soft modes. Explicitly minimizing over (unit) directors leads to an energy depending solely on deformation \cite{desimone2002macroscopic}
\begin{align}
W_I(\boldsymbol{\Lambda})  =\half \mu \left( \Lambda_1^2  \lambda_s + \Lambda_2^2 \lambda_s+ \Lambda_3^2/\lambda_s^2 \right), \mathrm{\ \ \ \ }\mathrm{Det}{\boldsymbol{\Lambda}}=1
\end{align}
where the ${\Lambda_i}$ are the ordered principle stretches. This form matches the \emph{ideal} LCE ``trace formula" (with $\lambda_s=r^{1/3}$), originally derived from  statistical mechanics \cite{bladon1993transitions, warner2007liquid}, and is minimized by any $\boldsymbol{\Lambda}_s(\n)$, revealing the  degenerate set of ground states (Fig.~\ref{fig:fig2}(a)). 
Furthermore, the energy is susceptible to microstructure (Fig.~\ref{fig:fig2}(a,b)). E.g., if an LCE cools forming $\n=\hat{\mathbf{y}}$ and is then stretched along $\hat{\mathbf{x}}$, then at an imposed stretch of $\lambda_s^{3/2}$, it will again attain a low energy state with $\n=\hat{\mathbf{x}}$. However, intermediate pure stretches are not soft, as the corresponding ground states require shear. Nevertheless, the LCE can accommodate pure stretch macroscopically by rapidly switching between bands of alternating shear:  a laminar ``stripe-domain" micro-structure \cite{finkelmann1997critical} that is geometrically compatible and soft.

More generally, any pure shear can be expressed as $\bfS(\boldsymbol{\gamma},\hat{\mathbf{m}})=\bfI+\boldsymbol{\gamma} \otimes\hat{\mathbf{m}}$, where  $\hat{\mathbf{m}}$ defines the shear plane and  $\boldsymbol{\gamma}$, orthogonal, encodes magnitude/direction. A lamination averaging to $\boldsymbol{\Lambda}$ is then constructed as:
\begin{align}
    &\boldsymbol{\Lambda}=f \boldsymbol{\Lambda}_1+ (1-f) \boldsymbol{\Lambda}_2 \mathrm{\ \ \ \ where} \\
    &\boldsymbol{\Lambda}_1= \bfS((1-f)\mathbf{a},\hat{\mathbf{m}}) \boldsymbol{\Lambda}, \mathrm{\ \ \ \ and \ \ \ \ } \boldsymbol{\Lambda}_2=\bfS(-f \mathbf{a},\hat{\mathbf{m}}) \boldsymbol{\Lambda}, \notag
\end{align}
and $0<f<1$ is the volume fraction of $\boldsymbol{\Lambda}_1$. If the resultant energy, $f W(\boldsymbol
{\Lambda}_1)+(1-f) W(\boldsymbol{\Lambda}_2)$, is lower than $W(\boldsymbol{\Lambda})$ then microstructure is favoured. The full relaxation of  $W_I(\boldsymbol{\Lambda})$ has  been constructed \cite{desimone2002macroscopic}, showing that any $\boldsymbol{\Lambda}$ with principle stretches less extreme than $\lambda_s$ may be achieved softly with double laminates. Since this set contains  $\boldsymbol{\Lambda}=\bfI$, the LCE can microstrucurally accommodate the experimental compression, motivating turbidity but not macroscopic surface instability.

The final missing physics is non-ideality: fabrication imprinted a preferred director $\n_0$, breaking degeneracy. Imprinting  generates an additional  bulk anchoring term (akin to $E$ or $B$ fields in liquid nematics \cite{ye2007semisoft}) parameterized  by a small coefficient $\alpha$ penalizing rotation away from $\n_0$  \cite{verwey1997compositional}:
\begin{align}
    W_{NI}(\boldsymbol{\Lambda}) =  \min_\n & \left( W_{NH}(\boldsymbol{\Lambda} \boldsymbol{\Lambda}_s^{-1}(\n)) \right. \\
& \left.+ \half \alpha \lambda_s \mu \mathrm{Tr}\left(\boldsymbol{\Lambda}(\bfI-\n_0\otimes\n_0) \boldsymbol{\Lambda}^\T \n\otimes\n \right)\right). \notag
\end{align}
We again minimize over $\n$ by collecting the relevant terms as $\half \mu \mathrm{Tr}(\bfM  \n \otimes \n)$, and directing $\n$ along $\bfM$'s  minimizing eigenvector. We then implement $W_{NI}$ in  finite elements to simulate the LCE slab (SI Sec. VII). An initial calculation clamping top and bottom surfaces ($\mathbf{\Lambda}=\mathbf{I}$) produces micro-structure suggestive of double lamination but with mesh-scale oscillations in director and deformation (Fig.\ S11). Releasing the top produces promising  thickness-scale topography, but atop  mesh-scale oscillations. Ultimately, this approach is unsatisfactory, as the  (non quasi-convex \cite{ball1976convexity, desimone2002macroscopic}) energy's minimizers are infinitely fine microstructures. Physically, resolution requires a Frank energy $\half K |\nabla \n|^2$ which smooths director variation over the ``nematic penetration depth''  $l \sim \sqrt{K/\mu} \sim 10^{-8}\mathrm{m}$ \cite{warner2007liquid}, regularizing the interfaces \cite{finkelmann1997critical}. Incorporating Frank energy enables converged mesh-independent results  \cite{zhou2021accelerated}, but requires meshes fine compared to $l$ that are infeasible for macroscopic samples. 

Instead, we adopt a two-scale approach, first computing the microstructural relaxation of $W_{NI}$ for any average deformation, then using this relaxed energy for macroscopic analysis. This relaxation problem is solved analytically for films in tension \cite{finkelmann1997critical, conti2002semisoft}, but a 3D result remains distant so we construct it numerically. The space of 3D deformations $\boldsymbol{\Lambda}$ is 9D, however incompressibility, frame indifference, and uniaxial material symmetry allow the energy to  be parameterized by just four scalar invariants  \cite{spencer1972deformations, destrade2013least}. We may thus re-express any nematic energy as $W(\boldsymbol{\Lambda})=w(\Lambda_1,\Lambda_3, \theta, \phi)$,  $\theta$, $\phi$ being the latitude and longitude of $\n_0$ in the (reference) stretch axes. To relax the energy,  we construct a grid of  $\sim$ 50k points over this space, spanning $1\le \Lambda_3\le 4$ with $\delta \Lambda_3 =0.1$ resolution, and  the remaining three (finite) dimensions with eleven even points each. We then compute a test deformation $\boldsymbol{\Lambda}_t(\Lambda_1,\Lambda_3, \theta, \phi)$ at each point, matching the invariants, and minimize over ``rank-1" laminates:
\begin{align}
W_{R1}(\boldsymbol{\Lambda})= \min_{f, \hat{\mathbf{m}}, \mathbf{a}\perp \hat{\mathbf{m}}} f W_{NI}(\boldsymbol{\Lambda}_1)+(1-f) W_{NI}(\boldsymbol{\Lambda}_2). 
\end{align}
\begin{figure}[ht]
	\includegraphics[width=\columnwidth]{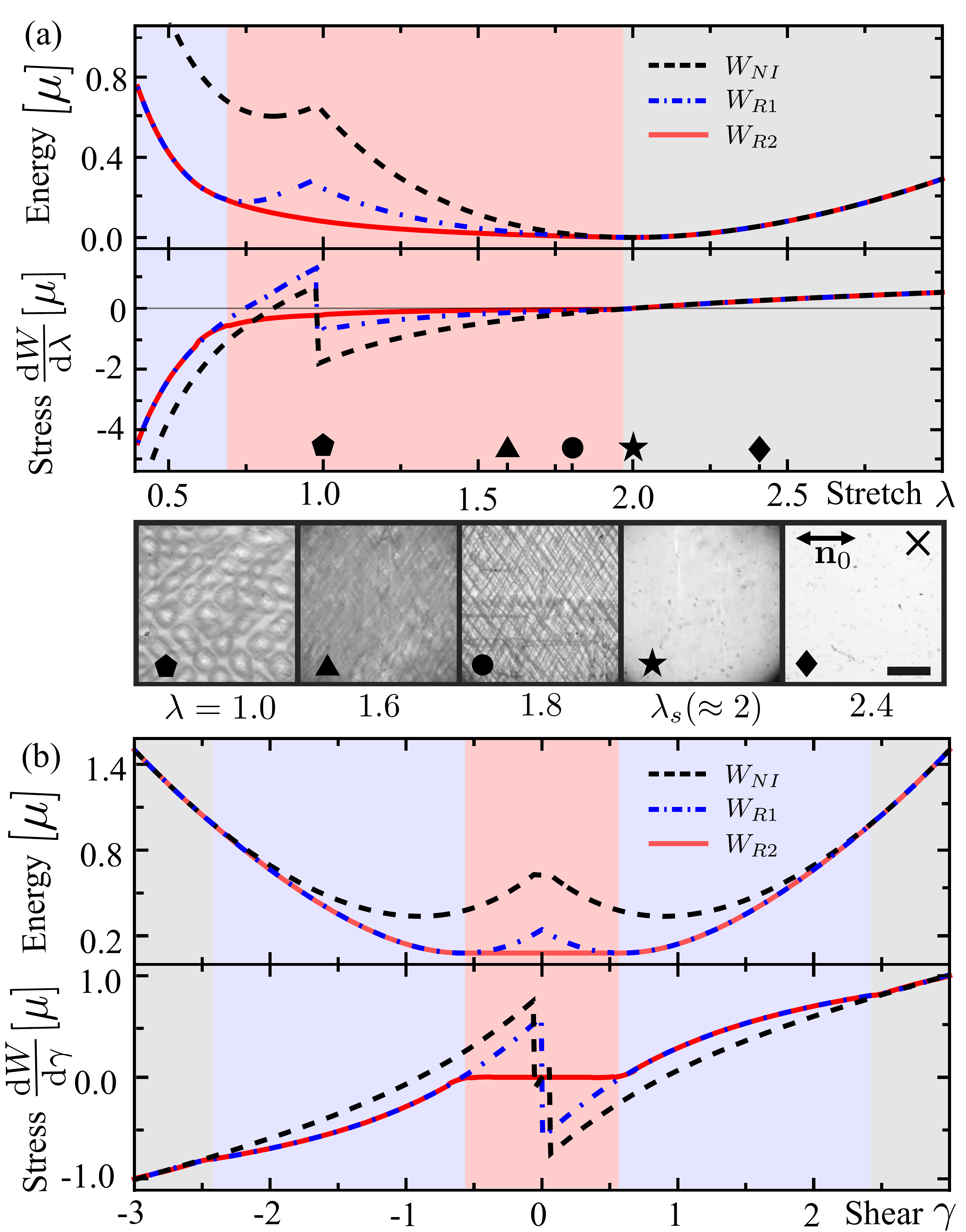}
	\caption{Energy and stress upon (a) uniaxial stretch $\boldsymbol{\Lambda}=\lambda \n_0 \otimes \n_0 +\lambda^{-1/2} (\bfI-\n_0\otimes \n_0)$ and (b) simple shear $\boldsymbol{\Lambda}=\bfI + \gamma (\bfn_0 \times \bfn_0^{\perp}) \otimes \bfn_0^{\perp}$. Gray domain: no micro-structure; blue: first-order laminate; red: second-order laminate. (a) bottom: POM of experimental samples ($t = 170 \mu m$, scale bar 500 $\mu m$) showing micro-structure evolution under uniaxial strain.  $\lambda=1$ shows surface instability, $1<\lambda<\lambda_s$ shows cross-hatch microstructure, and $\lambda \geq \lambda_s$ is a clear monodomain.}
	\label{fig:fig3}
\end{figure} 
Minimization over $f, \hat{\mathbf{m}}, \mathrm{\ and\ } \bfa$ (five d.o.f.) is conducted in Mathematica, using simulated annealing and conjugate gradient (SI Sec. VIII \cite{supp}). Interpolation over the grid then yields a numerical approximation of $W_{R1}$ and the lamination parameters. Repeating with $W_{R1}$ produces $W_{R2}$, while further iterations provide no benefit. For strong convergence, we conduct a final conjugate gradient minimization with analytic derivatives over the full 15 d.o.f. of double lamination. The relaxed first Piola-Kirchhoff stress for the $\boldsymbol{\Lambda}_t$ at each point is then be computed as the volume-weighted sum of the four constituent stresses, themselves evaluated as analytic derivatives  $P_{NI_{ij}}=\frac{\partial W_{NI}}{\partial \Lambda_{ij}}$ (SI Sec. VI \cite{supp}). Interpolation yields a numerical stress for any $\boldsymbol{\Lambda}_t$, which we rotate to produce the stress for any $\boldsymbol{\Lambda}$.

\begin{figure*}[ht]
\centering
	\includegraphics[width=\linewidth]{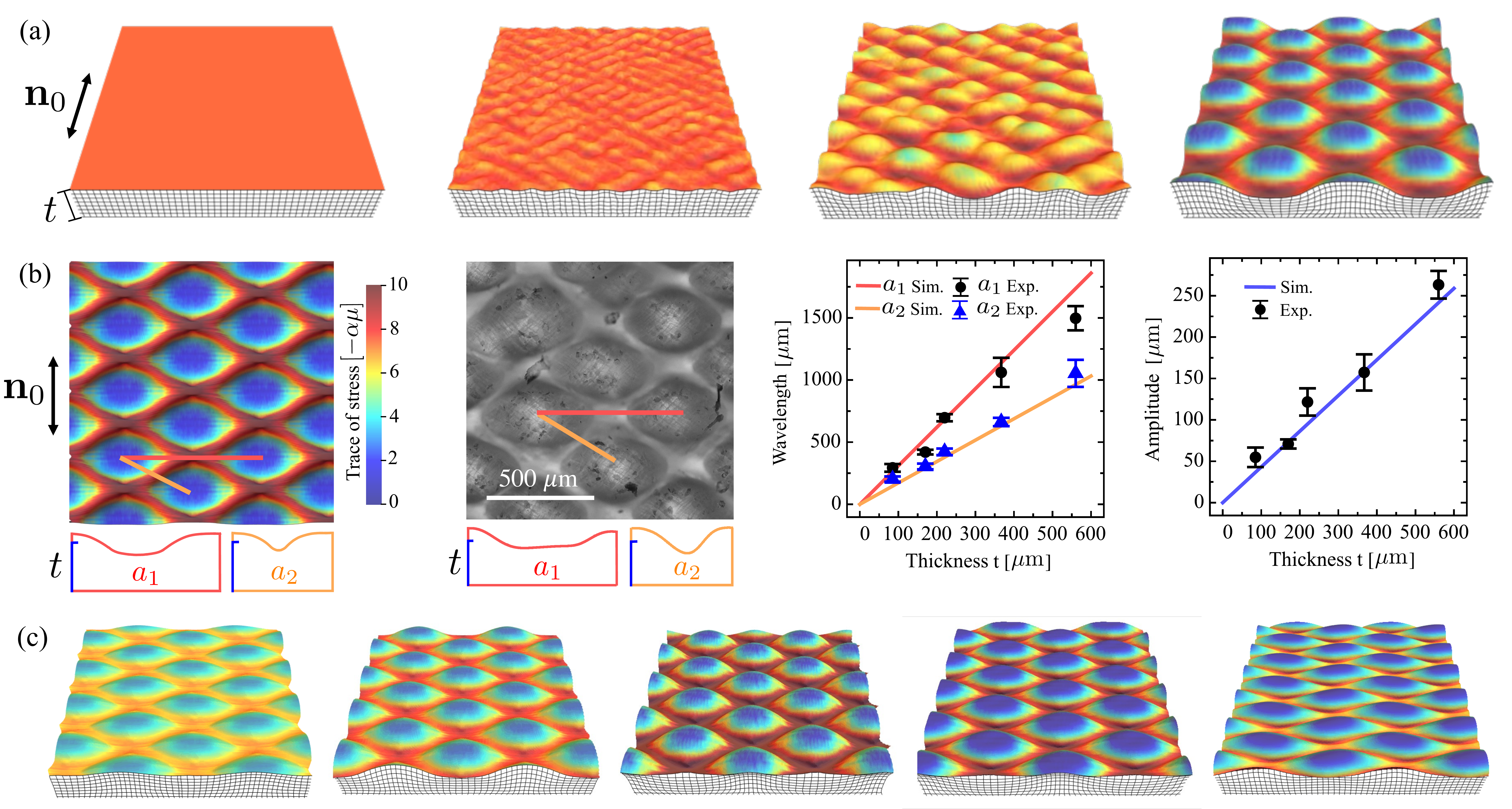}
	\caption{(a) Dynamics and coarsening of the surface instability with $(\lambda_s,\alpha) = (2,0.05)$. (b) Comparison of experimental ($t=220 \mu$m) and numerical $(\lambda_s,\alpha)=(2,0.05)$ topography. Wavelengths measured in POM, line profiles and amplitude from optical and scanning profilometry,  errors are  one sd, n = 10. (c) Simulated topography for $(\lambda_s,\alpha) = (1.44,0.05),~(1.71,0.05),~(2.15,0.05),~(2.15,0.15),~(2.15,0.3)$.}
	\label{fig:fig4}
\end{figure*} 

In Fig.\ \ref{fig:fig3} we show two resulting energy/force-extension curves: uniaxial stretch along $\bfn_0$ and simple shear perpendicularly. In both, microstructure occurs over a finite region, convexifies the energy, and eliminates discontinuities and negative stress gradients. Uniaxial stretching experiments (by adhering at intermediate temperatures) show a very fine micro-structure generating a hazy appearance in the expected region, which coarsens into a visible cross-hatch microstructure and then ultimately vanishes as the relaxed monodomain state, $\lambda_s$, is approached (Fig.\ \ref{fig:fig3}(a)). Samples adhered at $\lambda = 1$ also show the macroscopic surface instability, but these large features contain the fine microstructural cross-hatches within (Movie M4, Fig S8) justifying our two-scale approach. An additional test case, perpendicular stretching of a sheet (Fig.\ S13), agrees with  analytic results
\cite{finkelmann1997critical, conti2002semisoft}. Corresponding experiments (Fig.\ \ref{fig:fig2}b) confirm a stress plateau (and lamination) consistent with material values $\alpha \approx 0.05$, $\lambda_s \approx 2$.

Having tested the relaxed energy/stress functions, we minimize to predict the surface instability. We consider an LCE slab initially occupying $-t<z<0$, that is clamped below, free above and has planar $\n_0=\hat{\mathbf{x}}$. The final configuration minimizes the total energy $W = \int W_N(\boldsymbol{\Lambda}) \mathrm{d} V$ over permitted displacement fields $\mathbf{u}$, requiring stress balance, $\partial_i P_{ij}=0$, with $\bfP \cdot{\mathbf{\hat{z}}}=0$ on top. We first minimize using finite elements (SI Sec. IV and Sec. X). For simplicity, we mimic incompressibility by taking $W_N$ as compressible but with a large bulk modulus of $B=100\mu$:
\begin{align}
W_N(\boldsymbol{\Lambda})=W_{R2}(\boldsymbol{\Lambda}/\mathrm{Det}{\boldsymbol{\Lambda}}^{1/3})+\half B (\mathrm{Det}{\boldsymbol{\Lambda}}-1)^2.
\end{align}
 An initial computation considers a near square slab  ($6.24t\times 6 t$)  divided into 126k elements of size $0.06 t$.
The LCE is modeled with $\lambda_s=2$, $\alpha=0.05$, and initialized at $\boldsymbol{\Lambda}=\bfI$. As seen in Fig.\ \ref{fig:fig4}(a) and SI movie M5 \cite{supp}, the free surface immediately destabilizes to oblique mesh-scale ripples, which coarsen via dynamics into an equilibrium cross-hatch strongly resembling experiment. We then use fine-mesh calculations on geometrically optimized unit cells to find the exact minimizers for a range of $\lambda_s$ and $\alpha$ (i.e.\ different LCEs). All are unstable, but with lower amplitude at lower $\lambda_s$, and shifting wave-vectors at higher $\alpha$ (Fig.~\ref{fig:fig4}(c)). The range of $\alpha$ explored spans the reported range for physical monodomain LCEs \cite{finkelmann1997critical}, suggesting the cross-hatch instability is ubiquitous in  LCEs under compression.
Equilibrium amplitude and wavelength are necessarily simply proportional to the problem's only length-scale, $t$, and in good agreement with experiment. Moreover, unlike creasing, all topography is smooth and non-contacting.

Linear stability analysis offers further insight. Owing to finite compressibility, the equilibrium equations  admit a transitionally invariant base state $\boldsymbol{\Lambda}_0 = \bfI + \gamma  \hat{\mathbf{z}}\otimes\hat{\mathbf{z}}$, where $\gamma \sim \alpha \mu/B \sim 10^{-5}$. Following Biot \cite{biot1965mechanics}, we consider an infinite depth slab (half-space), and add an small incremental displacement, $\mathbf{u}= \gamma \hat{\mathbf{z}} + \epsilon \mathbf{u}_1$, giving $\boldsymbol{\Lambda}=\boldsymbol{\Lambda}^0+\epsilon \boldsymbol{\Lambda}^1 $. Since perturbations are about equilibrium, energy varies quadratically
\begin{align}
W_N =W_N(\boldsymbol{\Lambda}^0)+ \frac{1}{2} \epsilon^2 \frac{\partial^2 W_N}{\partial \Lambda_{ij} \partial \Lambda_{lm}} \bigg|_{\boldsymbol{\Lambda}^0}  \Lambda^1_{ij}  \Lambda^1_{lm}.
\end{align}
Minimizing variationally over displacement gives the  incremental equilibrium equations, $ \partial_i  P^{1}_{i j} =0$ and $\bfP^{1} \hat{\mathbf{z}} =0$, where  $P^{1}_{ij}=\frac{\partial^2 W_N}{\partial \Lambda_{ij} \partial \Lambda_{lm}} \big|_{\boldsymbol{\Lambda}_0}  {\Lambda}^1_{lm}$. Many of these derivatives are zero via uniaxial material symmetry. The remainder, we compute using finite differences, reducing the energy to a quadratic in $\nabla \mathbf{u}_1$, and the stress-equations to  linear constant-coefficient ODEs. To compute accurate derivatives, we explicitly re-minimize $W_{R2}$ at the finite-difference points, revealing additional zero curvatures corresponding to vanishing incremental moduli for shearing perpendicular to $\n_0$ (c.f.\ the $W_{R2}$ plateau in Fig.~\ref{fig:fig3}(b)). Such vanishing moduli have been observed at the onset of monodomain striping \cite{ye2007semisoft, biggins2008semisoft} but are here deep within the microstructure region. The underlying cause is that optimal lamination involves four equivalent deformations, and such shears are accommodated by rearranging volume fractions, similar to how, in 1D, curvature vanishes after common-tangent convexification.

\begin{figure*}[!ht]
\centering
	\includegraphics[width=\linewidth]{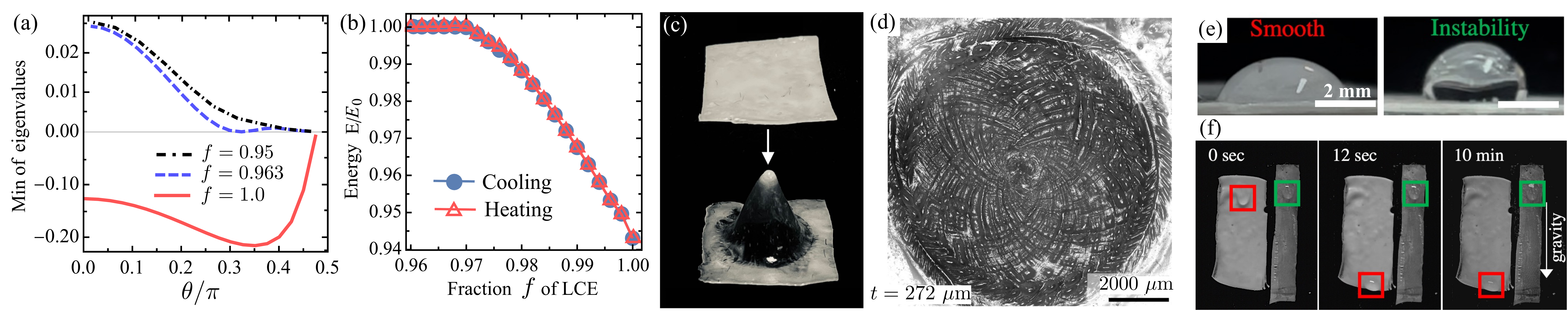}
	\caption{(a-b) Analysis of onset. (a) Minimal Hessian eigenvalue vs angle for $f=0.95$ (stable), $f=f_c=0.963$ (instability threshold) and $f=1.0$ (pure LCE, unstable). For $f=1$, the minimum  eigenvalue has $\theta_c\approx0.35\pi$. (b) Cooling and heating FEM showing super-critical onset at $f_c \approx 0.97$. $E_0$ is the base-state energy. (c-f) Experimental extensions. (c) Mechanically programmed LCE cone. (d) Micrograph of resulting log-spiral surface instability. (e) Photographs of a 15 $\mu l$ water droplet on room temperature LCEs ($t=170\mu m$) with smooth and unstable surfaces: instability  changes the contact angle. (f) Similar  droplet running down vertical surfaces: instability promotes pinning.}
	\label{fig:fig5}
\end{figure*} 
To solve, we substitute an ansatz that decays with depth,  $\mathbf{u}_1 =\exp(k \lambda z)\mathbf{v}(x,y)$, and undulates in plane with $k$ vector along $\mathbf{x}'$, forming an angle $\theta$ with $\mathbf{x}$: 
\begin{equation}
\mathbf{v}= c_h\cos(k x') \hat{\mathbf{x}}'  +  c_g \cos(k x') \hat{\mathbf{y}}'+ \sin(k x') \hat{\mathbf{z}}. \nonumber
\end{equation}
Substituting into the stress equations (SI Sec. XI \cite{supp})  trigonometric, exponential and $k$ factors all cancel, leaving dimensionless algebraic equations for $c_g,$  $c_h$ and $\lambda$. Solving for $c_g(\lambda)$ and $c_h(\lambda)$ reveals a final  quadratic in $\lambda^2$ giving two decaying roots ($\lambda_1, \lambda_2$), and hence:
\[
\mathbf{u}_1=c_1 \exp(k \lambda_1 z) \mathbf{v}(\lambda_1)+c_2 \exp(k \lambda_2 z) \mathbf{v}(\lambda_2).
\]
We then substitute $\mathbf{u}_1$ into the energy and integrate,  yielding a quadratic in $c_1,\mathrm{\ }c_2$. Stability follows from the Hessian, with negative eigenvalues indicating growing perturbations. Inevitably given the scale-free character, stability is independent of $k$. More surprisingly, despite uni-axial symmetry, all angles are unstable. We thus assess relative growth rates from the eigenvalue's magnitude, revealing an oblique perturbation grows fastest (Fig.\ \ref{fig:fig5}(a)), motivating the cross-hatch.

To study onset, we imitate a cooling LCE via the energy  $(1-f) W_{NH}(\boldsymbol{\Lambda})+f W_{R2}(\boldsymbol{\Lambda})$, with $f$ a temperature proxy. Both finite-elements and linear analysis show onset around $f\approx0.96$. Furthermore, finite elements show onset is super-critical, with topography appearing continuously and without hysteresis (Fig.~\ref{fig:fig5}(b)), unlike the strong subcriticality of Biot creasing. 


 Our work thus shows that the  exotic constitutive law  of LCEs  still supports a compressive surface instability, with similar scale-free properties to Biot creasing, but also fundamental differences in pattern, criticality and singularity. We anticipate similar behavior will be found in other soft solids with distinctive constitutive laws: highly anisotropic solids (fiber reinforced elastomers, biological tissues), phase changing solids, active nematic solids and mechanistic meta-materials are all important and analogous examples. Hopefully, such studies will also generate a stronger intuition for Biot-type instabilities, and ultimately enable a predictive analytic framework. Within LCEs, our rigorously homogenized constitutive law also enables quantitative mechanical engineering including studies of other instabilities  \cite{mihai2021instabilities, giudici2020giant, lee2021actuation,lee2023universal,krieger2019tunable, goriely2021liquid},  and our homogenization approach will similarly enable analysis of other phase-changing materials \cite{bhattacharya2003microstructure, adams2007soft, gu2021exploding}.

The LCE surface instability is also an exemplar of instabilities providing complex morphing without correspondingly complex fabrication. A limitation is that this approach provides little control over pattern. However, some more complex morphing may be regained by combining instability with simple patterned fabrication. As an example, we repeat our experiment, using an LCE slab containing a radial director pattern (fabricated by forming a cone during programming), which generates cross-hatches in a striking log-spiral (Fig.\ \ref{fig:fig5}(c,d)). 
The instability's switchable high amplitude topography suggests applications in smart surfaces. For example,  a droplet sitting atop the instability (Fig.\ \ref{fig:fig5}(e,f)) forms a Wenzel state \cite{wenzel1936resistance,lafuma2003superhydrophobic} with both  a  higher contact angle and stronger pinning than one atop  an identical smooth LCE. Such hydrophobic pinning is observed on rose petals, which have a strikingly similar topography \cite{Roy2019,Oopath2022a}. Further smart-surface applications could include switchable aerodynamics, haptics, adhesion and friction.

\begin{acknowledgments}
	This work was supported by a UKRI ‘future leaders fellowship’ grant (grant no. MR/S017186/1).
\end{acknowledgments}
\bibliography{elastomer}
\end{document}


\title{Supplemental Information: \\Surface instability in a nematic elastomer}

\author{Morgan Barnes}
\author{Fan Feng}
\author{John S. Biggins}
\email{jsb56@cam.ac.uk}
\affiliation{Department of Engineering, University of Cambridge, Trumpington St., Cambridge CB2 1PZ, United Kingdom}

\date{\today}

\maketitle


\setcounter{equation}{0}
\setcounter{figure}{0}
\setcounter{table}{0}
\setcounter{page}{1}
\setcounter{section}{0}
\makeatletter
\renewcommand{\theequation}{S\arabic{equation}}
\renewcommand{\thefigure}{S\arabic{figure}}
\renewcommand{\thetable}{S\arabic{table}}
\renewcommand{\bibnumfmt}[1]{[S#1]}
\renewcommand{\citenumfont}[1]{S#1}


\section{Material Fabrication}

\subsection{Materials}
 1,4-bis-[4-(3-acryloyloxypropyloxy)benzoyloxy]-2-methylbenzene (RM257) was purchased from Wilshire technologies. 2,20
-(Ethylenedioxy) diethanethiol (EDDET), pentaerythritol tetrakis
(3-mercaptopropionate) (PETMP), chloroform, dipropyl
amine (DPA), and (2-hydroxyethoxy)-2-
methylpropiophenone (HHMP) were obtained from Sigma Aldrich. All materials were used as received.

\subsection{Synthesis}
LCEs were synthesized as previously reported. \cite{barnes2019direct} RM257 (588 mg, 1 mmol), EDDET (124.09 mg, 0.6818 mmol), PETMP (55.45 mg, 0.1136 mmol), 0.5 wt\% HHMP (3.84 mg), and 40 wt\% chlorofrom (307 mg) were combined in a scintillation vial, heated to ~80$^o$C, and vortexed until all reagents were dissolved and mixed. Next, 0.23 mol\% of DPA  diluted to 2 wt\% in chloroform (0.417 mg DPA in 20.44 mg chloroform) was added to the mixture to catalyze the thiol-acrylate Michael addition. The mixture was deposited between two glass slides separated by spacers of desired thickness. After 2 hours the LCE film was removed from the glass cell and heated to 100$^\circ$C for 1 hour to evaporate the remaining chloroform, resulting in an opaque polydomain LCE. For monodomain alignment, the LCE was stretched to double its length and UV cured for 10 minutes to crosslink the excess acrylates. The resulting spontaneous actuation, $\lambda_s$ is shown in Figure \ref{fig:actuation_v_temperature}. For a cone LCE, the polydomain LCE was clamped to metal sheet with a circular cutout of the desired cone diameter. A toothpick was then inserted into the center of the circle to produce a cone, the toothpick was removed to allow the cone to relax to $\lambda\approx 2$ and then UV cured. 

\begin{figure}[ht]
\includegraphics[width=1\textwidth]{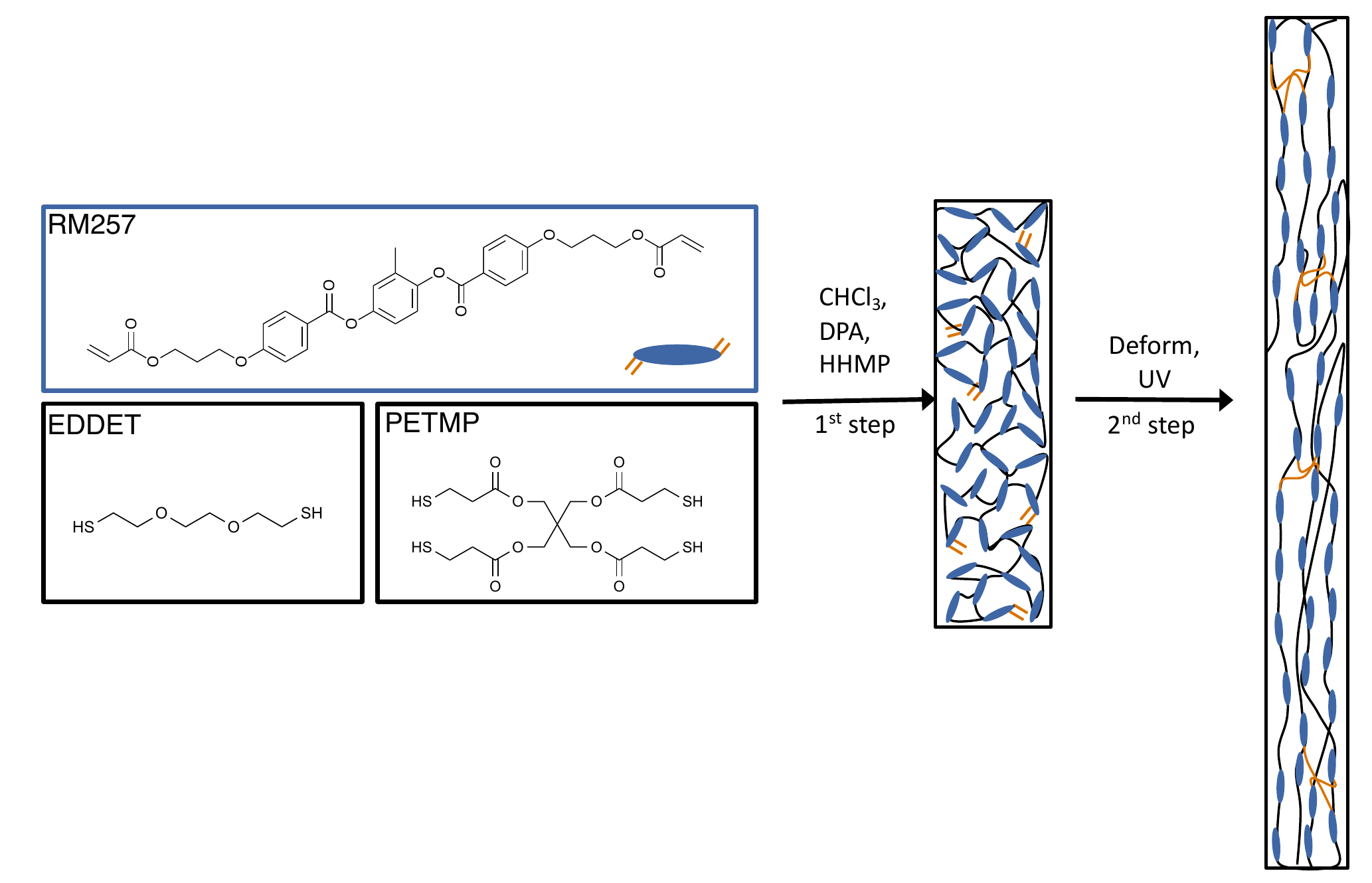}
	\caption{Schematic of LCE chemistry and two-step synthesis.}
	\label{fig:chemistry}
\end{figure}

\subsection{Surface Instabilities}
Surface instabilities for strains  = $\lambda_s$ were achieved by heating programmed LCEs to their fully actuated (isotropic) flat state (130$^o$C) and quickly pressing a glass slide coated with brush-on Loctite super glue onto the LCE. After adhered the LCE is cooled to room temperature to produce the surface instabilities (Figs. 1(b), 4(b), 6, \ref{fig:spiral})  

\subsection{Crosshatch Microstructure}
To achieve cross-hatch microstructures for strains less than $\lambda_s$, (as shown in Fig. 3) the LCEs were uniformly heated to temperatures less than 130$^\circ$C to achieve partial contraction, adhered onto a glass slide, and cooled to room temperature. 

\section{General Characterization}

\subsection{Tensile Testing}
Tensile data was acquired using an Instron 5584 equipped with a 10 N load cell. An LCE (10 mm X 0.23 mm X 30 mm) was stretched perpendicularly to the programmed director at a pulling speed of 0.0009 $\%$ strain/second.

\subsection{Polarized Optical Microscopy}
Cross-polarized optical micrographs were taken using a Nikon Eclipse LV100nd microscope in transmission mode. 

\subsection{Actuation}
\begin{figure}[ht]
\includegraphics[width=0.5\textwidth]{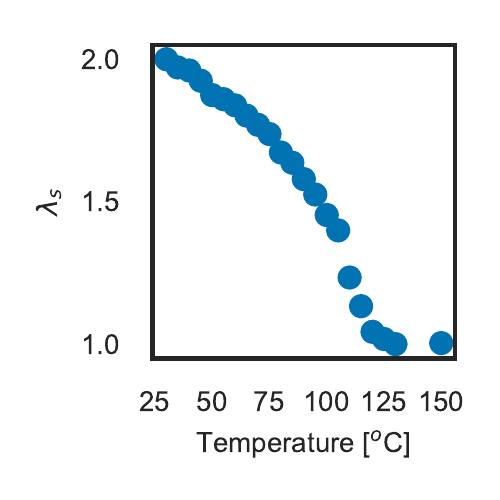}
	\caption{Spontaneous actuation of a monodomain LCE in response to temperature.}
	\label{fig:actuation_v_temperature}
\end{figure}

Actuation vs. temperature (Fig.~\ref{fig:actuation_v_temperature}) data was acquired by measuring the distance between points of a monodomain LCE while heating the LCE using a Linkham LTS420 hot stage from room temperature to $150^\circ$C with a temperature step of 5$^\circ$C and allowing the LCE to equilibrate for 5 minutes after each step. 

\subsection{Water contact angle and pinning}
Qualitative water contact angle and pinning was compared in a smooth LCE, which was UV cured at a 0 strain state and used in the equilibrium polydomain state, and an LCE with surface instability ($t = 170 \mu m$, strained to 100\% and glued onto a glass slide at $T = 130 ^o$C).  15 $\mu$l water droplets were deposited on both LCEs at room temperature and immediately imaged from the side to observe the contact angles. The LCEs were then rotated 90$^o$C and videoed to track the time it took the droplet to roll off the smooth LCE and the water pinning effect in the surface instability LCE. 
\section{Surface Instability Characterization}

\subsection{Optical Profilometry}

Depth topography images were obtained with an Olympus BX51 optical microscope. Montage images (Fig.~\ref{fig:montage_micrographs}) and height maps (Fig.~\ref{fig:depth_micrographs}) were created within the Olympus LAS software by merging the z-stack images to calculate the optimum focal depth for each pixel on the surface. Wavelength measurements were calculated by manually calculating the periodic distance between surface features in the montage images. Amplitude measurements were calculated by extracting the local minimum and maximum for each surface feature period. 

\begin{figure}[!ht]
\includegraphics[width=\columnwidth]{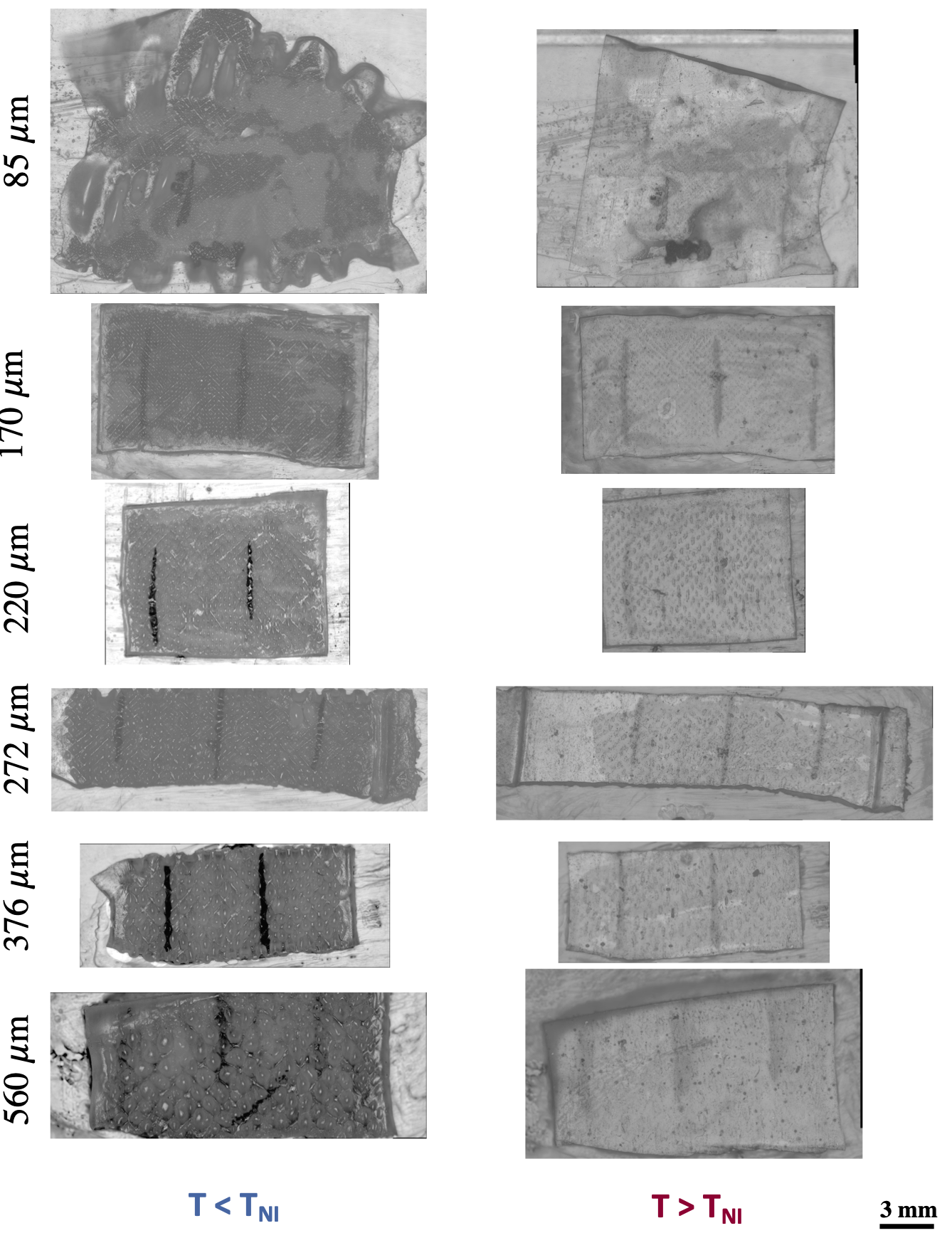}
	\caption{Micrographs of surface instability for monodomain LCEs with varying thicknesses when hot (130$^\circ$C) and cold (room temperature). Heated LCEs show some optical artifacts of surface instability but the surface is smooth as shown in the heated depth topography as shown in Fig. 1(b).}
	\label{fig:full_samples}
\end{figure}

\begin{figure}[!ht]
\includegraphics[width=\columnwidth]{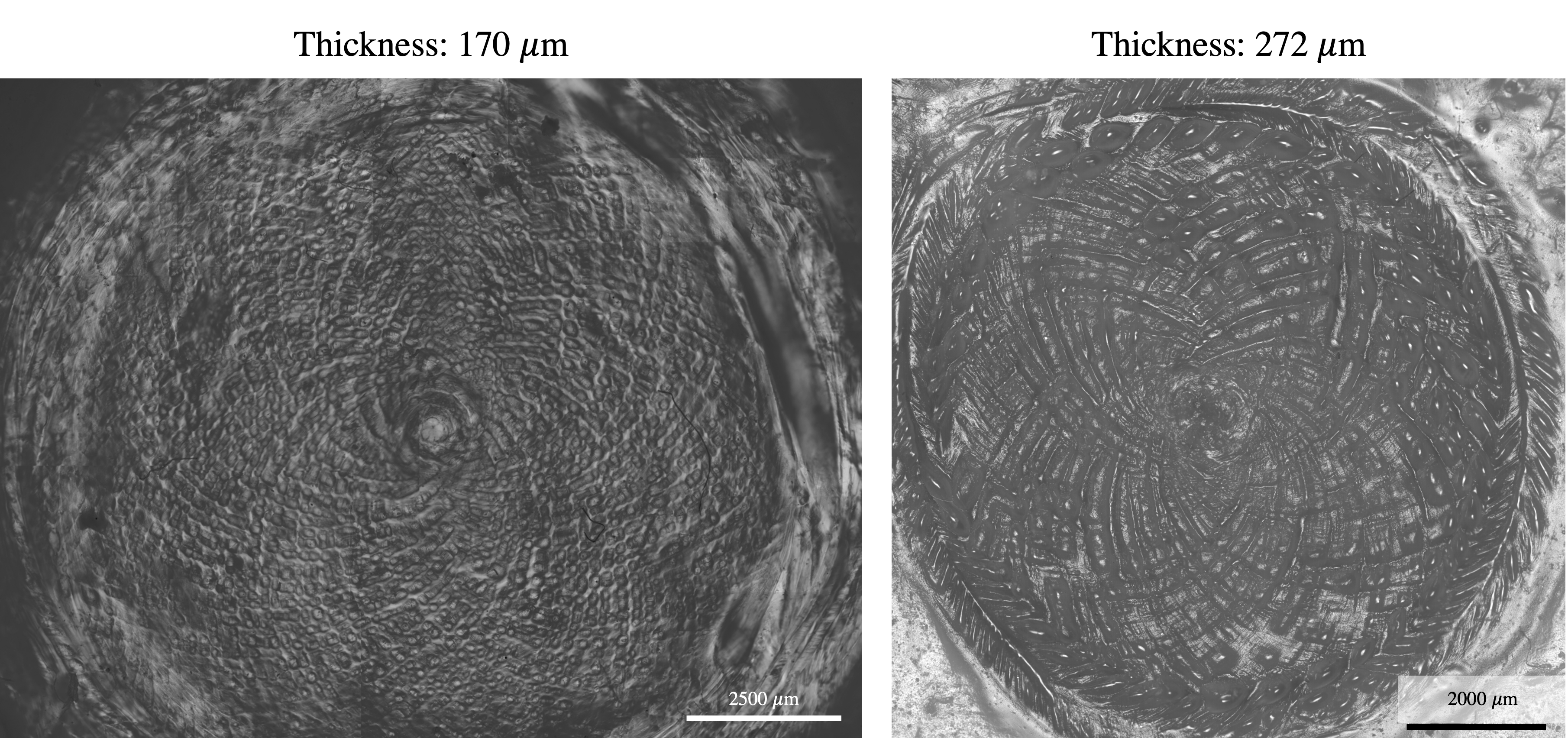}
	\caption{Micrographs of the surface instability of LCE cones adhered to glass at $T=130^\circ$C.}
	\label{fig:spiral}
\end{figure}

\begin{figure}[!ht]
\includegraphics[width=\columnwidth]{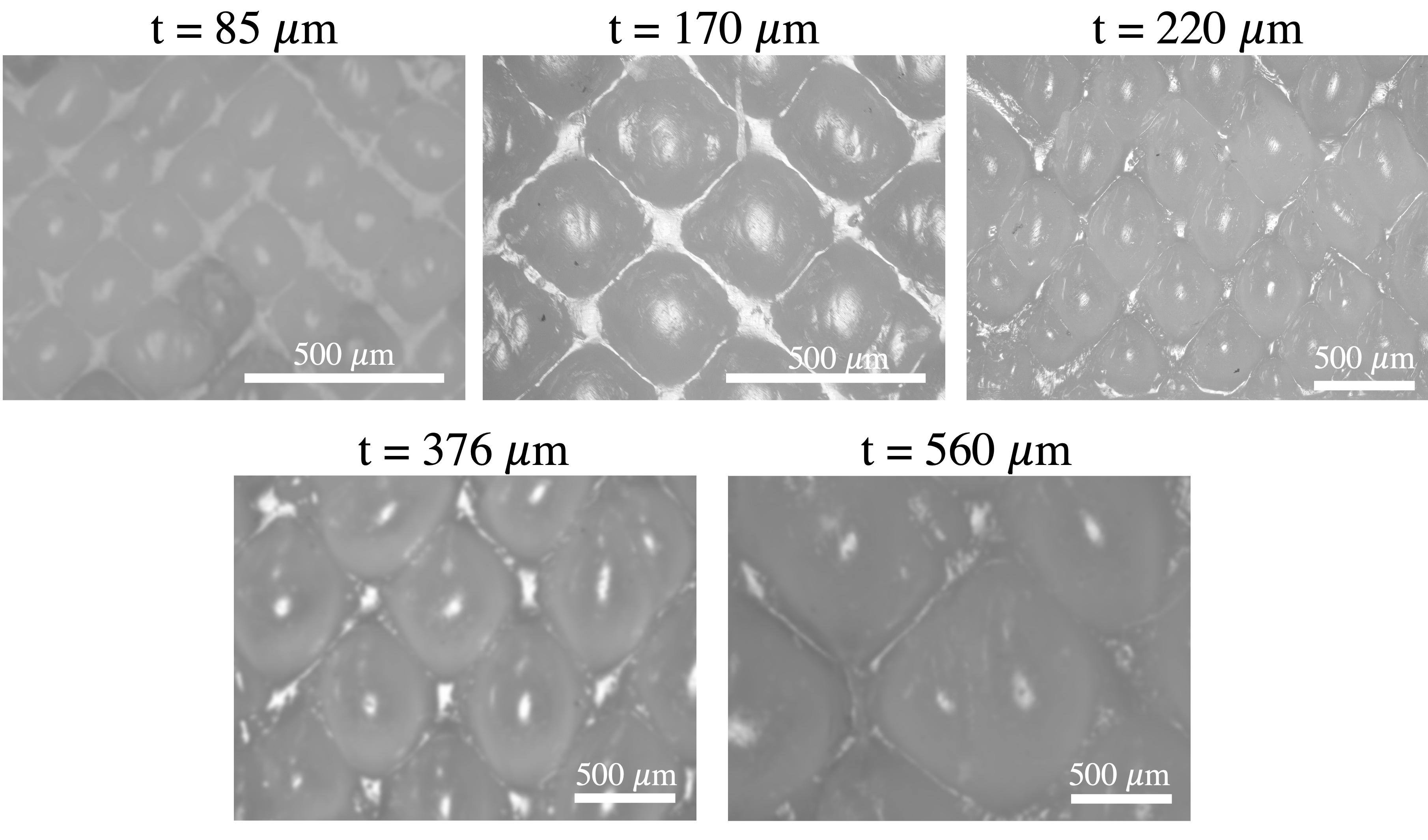}
	\caption{Micrographs of surface instability of monodomain LCEs with $\lambda_s$ = 2 and glued down at the fully contracted ($\lambda$=1) state at varying thicknesses.}
	\label{fig:montage_micrographs}
\end{figure}

\begin{figure}[!ht]
\includegraphics[width=\columnwidth]{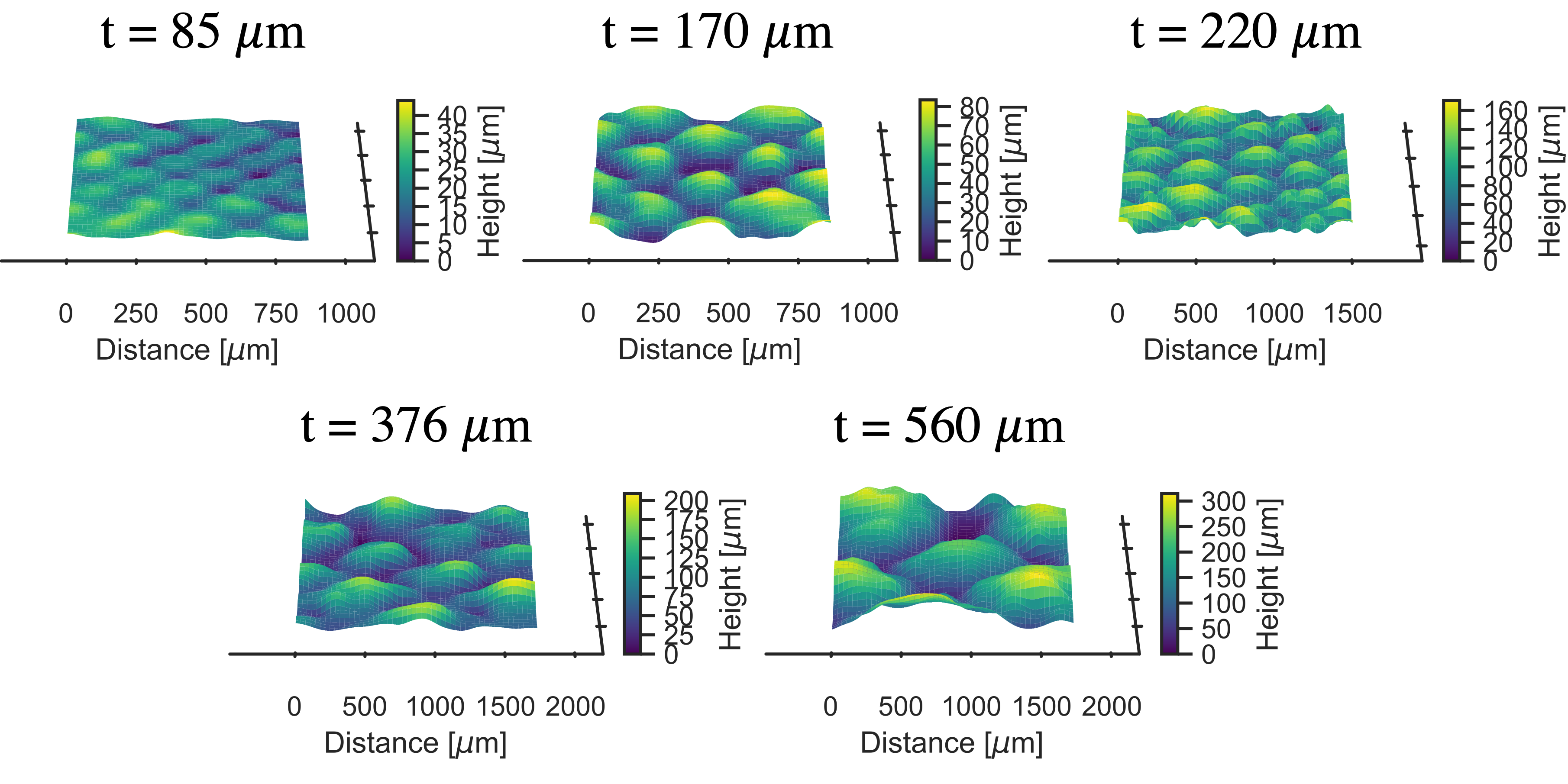}
	\caption{Depth topography of the surface instability of monodomain LCEs with $\lambda_s$ = 2 and glued down at the fully contracted ($\lambda$=1) state at varying thickness.}
	\label{fig:depth_micrographs}
\end{figure}

Large length scale images (Figs. \ref{fig:full_samples}, \ref{fig:spiral}) were captured using the stitching feature on the LAS software with 20\% overlap.

\subsection{Scanning Profilometry}
\begin{figure}[!ht]
\includegraphics[width=\columnwidth]{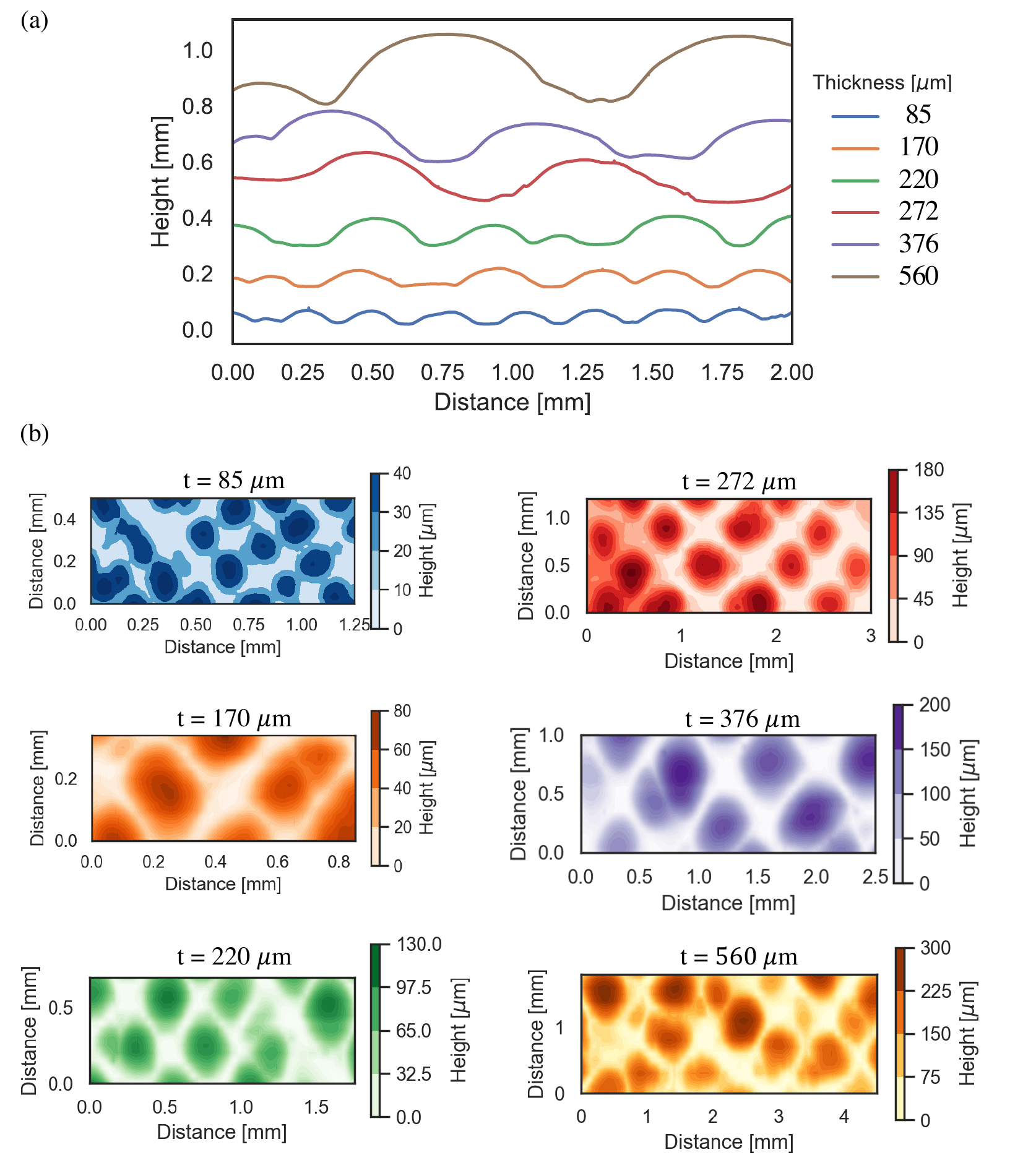}
	\caption{(a) Representative line scans (with vertical line shifts for viewability) of surface topography of monodomain LCEs with $\lambda_s$ = 2 and glued down at the fully contracted ($\lambda$=1) state at varying thickness. (b) Constructed height contour maps created from a series of line scans.}
	\label{fig:line_scans}
\end{figure}

Height line scans (Fig.~\ref{fig:line_scans}(a)) were collected using a Taylor Hobson Talysurf 120 profilometer and scanned parallel to the nematic director. Surface depth contour plots were created by stacking multiple line scans (Fig.~\ref{fig:line_scans}(b)) and amplitudes of the surface features were calculated by extracting the local minimum and maximum for each surface feature period. 

\begin{figure}[!ht]
\includegraphics[width=\columnwidth]{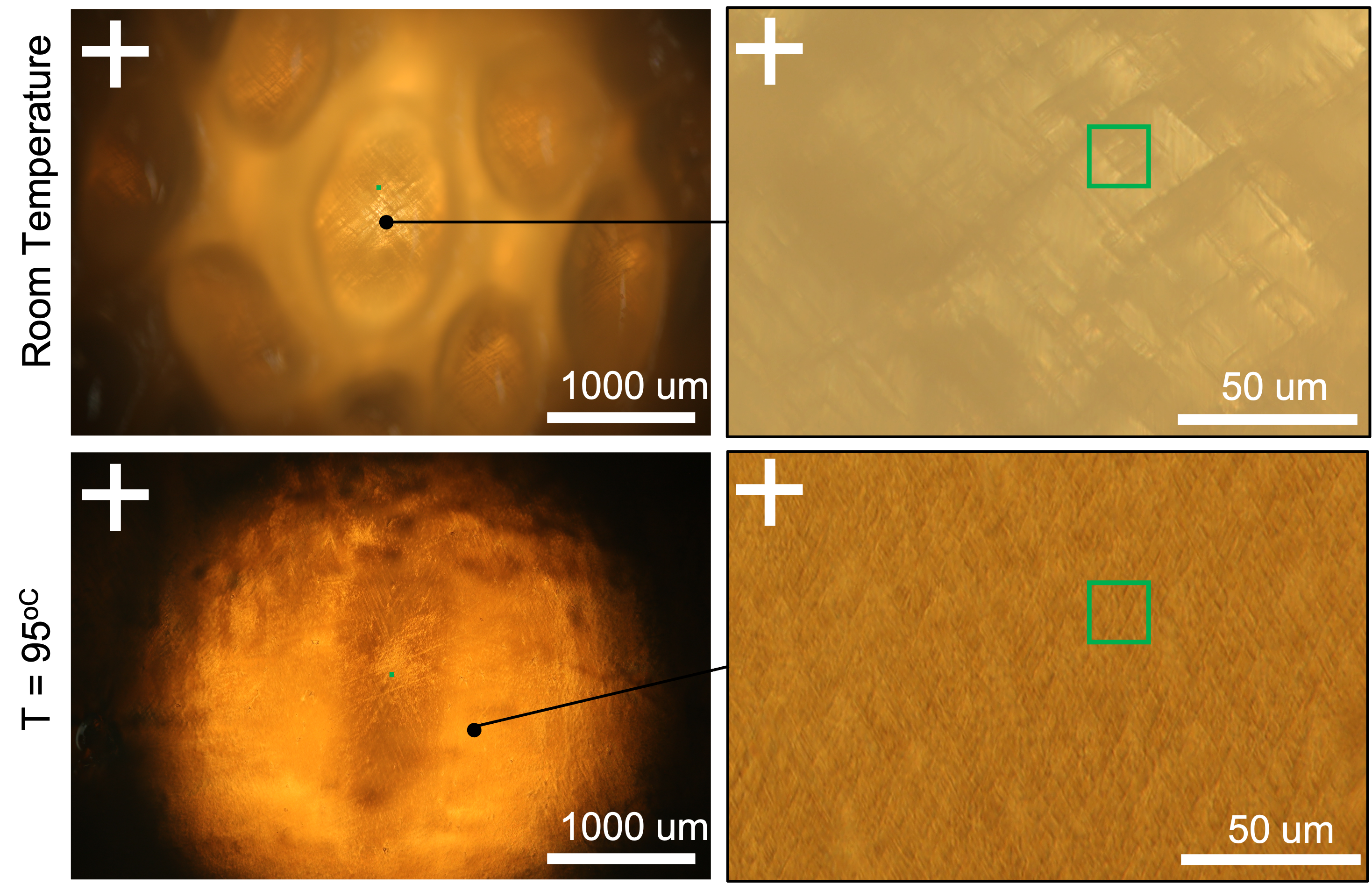}
	\caption{Cross-polarized micrographs of small-scale microstructure within a 560 um thick film undergoing the surface instability. At room temperature  the macroscopic surface instability (top-left, ~1000um wavelength) evidently also contains small-scale cross-hatch micro-structure within it. In the proud domed regions, partial strain relief and the proximity of a free surface yield comparatively coarse microstructure (top right), while the valley regions are hazy indicating fine light-scattering microstructure that cannot be imaged at room temperature.  However, we are able to visualize this finer microstructure by heating the sample just above the nematic-isotropic transition (bottom left, same region as top left), which dramatically reduces the LCE's optical anisotropy and hence its light scattering scattering. Magnifying a valley region (bottom right) reveals a fine microstructural crosshatch at ~4 um or 0.714\% of the thickness.  The size of a characteristic finite-element used in simulations for a 560 um thick film is illustrated by the green box (16.5 $\mu$m $\times$ 16.5 $\mu$m), which is much smaller than the feature size, but larger than the microstructure, justifying the two scale approach.
}
	\label{fig:microstructure}
\end{figure} 

\clearpage

\newpage
\section{Overview of hyper-elastic modeling of the LCE surface instability}
As introduced in the main text,  we model the LCE as an slab that is infinite in the $x-y$ plane, and initially occupies $-t<z<0$. Upon cooling from isotropic to nematic, the slab undergoes a displacement field $\mathbf{u}$ and resulting deformation gradient $\boldsymbol{\Lambda}=\bfI+\nabla \mathbf{u}$. The bottom surface ($z=-t$) is clamped ($\mathbf{u}=0$) while the top surface is free. We model the LCE as a hyper-elastic solid, with a stored energy density $W(\boldsymbol{\Lambda})$, and hence a total energy 
\begin{equation}
E=\int_{-\infty} ^\infty \int_{-\infty} ^\infty \int_{-t} ^{0} W(\boldsymbol{\Lambda}) \mathrm{d}x \mathrm{d}y \mathrm{d}z. \label{eq:totalenergy}
\end{equation}
The configuration of the LCE is given my minimizing this  energy variationally over  displacements $\mathbf{u}$, leading to the expected equilibrium equations and boundary conditions:
\begin{equation}
\nabla \cdot \mathbf{P} =\mathbf{0},\mathrm{\ \ \ \ \ \ \ }\mathbf{P}\cdot \hat{\bfz}\big|_{z=0}=\mathbf{0}\mathrm{,\ \ \ \ \ \ \ }\mathbf{u}\big|_{z=-t}=0,
\end{equation}
where 
\begin{equation}
P_{ij}=\frac{\partial W}{\partial \Lambda_{i j}}
\end{equation}
is the first Piola–Kirchhoff (PK1) large-deformation stress tensor. The hyper-elastic laws we consider are formulated for strictly incompressible LCEs, and hence subject to the constraint $\mathrm{Det}(\boldsymbol{\Lambda})=1$. In analytic work, this is typically implemented by adding a pressure Lagrange multiplier term $p (\mathrm{Det}(\boldsymbol{\Lambda})-1)$ to $W$. However, for our numerical work, we instead replace the strictly incompressible energies by counterparts with a large but finite bulk modulus $B$,
\begin{align}
W(\boldsymbol{\Lambda}) \to W(\boldsymbol{\Lambda}/\mathrm{Det}{\boldsymbol{\Lambda}}^{1/3})+\frac{1}{2} B (\mathrm{Det}{\boldsymbol{\Lambda}}-1)^2. \label{eq:SI_compressible}
\end{align}

\subsection{Finite element minimization}\label{sec:fem}

\begin{figure}[h]
	\includegraphics[width=\columnwidth]{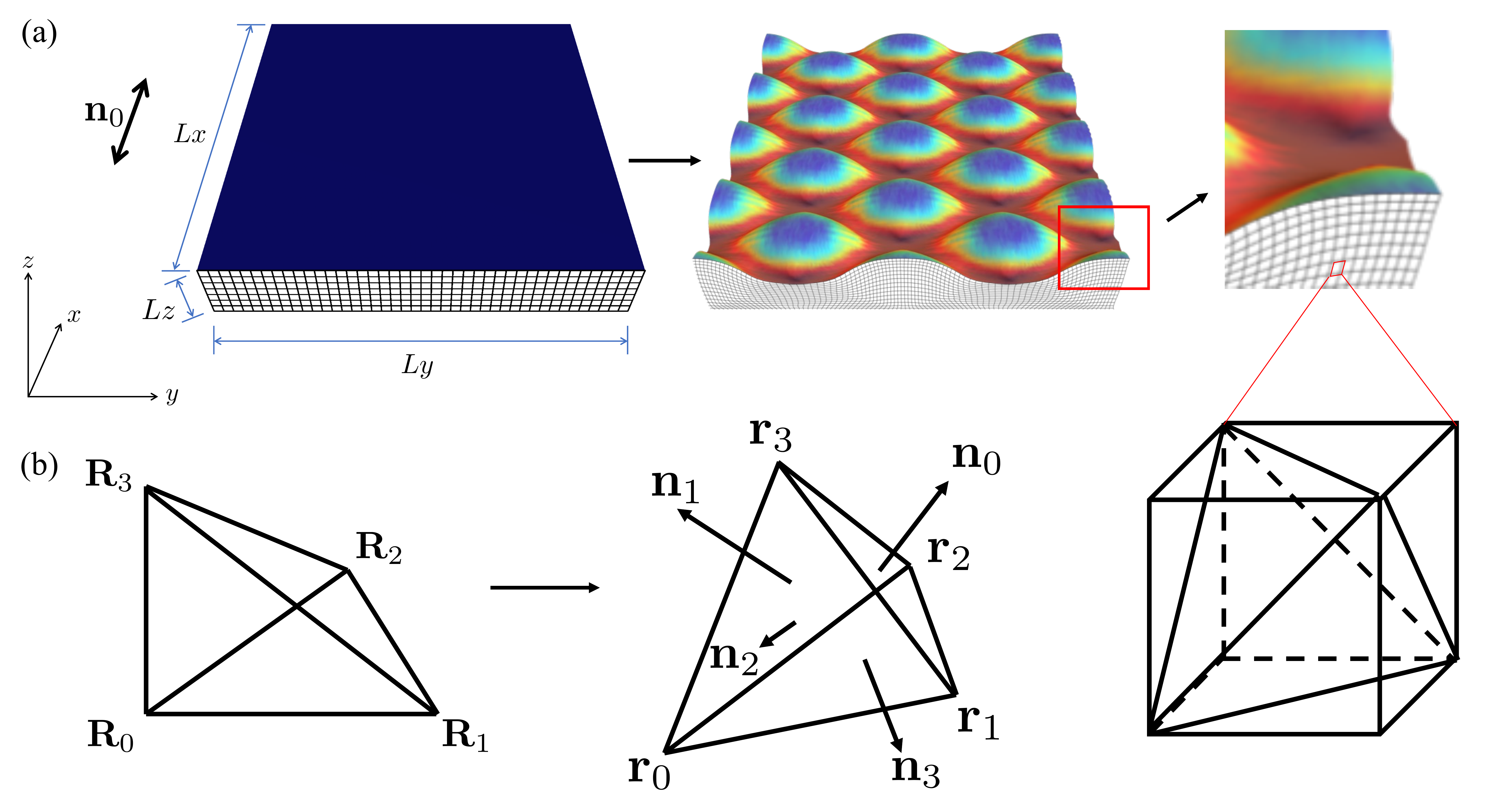}
	\caption{(a) The simulation is initialized for an LCE slab at the isotropic state. As cooling to the nematic state, the surface instability is triggered and simulated by FEM. Each cuboid element has nine tetrahedrons. (b) The reference and deformed tetrahedrons and the surface normals.}
	\label{fig:finite_element}
\end{figure} 
To explore the behavior of different constitutive laws, we repeatedly minimize the total energy (\ref{eq:totalenergy}) using finite elements. Our finite elements are based on the bespoke C code in \cite{tallinen2013surface}, but, for completeness we give a brief overview of the method here. The code considers a finite rectilinear region of LCE layer, with dimensions $L_x \times L_y \times L_z$ as depicted in Fig.~\ref{fig:finite_element}(a). The slab has a clamped base, free top surface, and periodicity imposed in-plane. The preferred director $\bfn_0$ is always aligned along $x$. The layer is discretized into a cuboidal mesh of nodes with number  $n_x$, $n_y$, and $n_z$ in each dimension. The $i$th node will move during the calculation from its reference positions $\mathbf{R}_i$ to final positions $\mathbf{r}_i=\mathbf{R}_i+\mathbf{u}_i$, in order to find the energy minimizing configuration.

To compute and minimize the elastic energy, each cuboid is then divided into five tetrahedra which  form constant strain finite elements.  More precisely, the uniform deformation gradient in each tetrahedra is  simply given by $\bflambda = (\bfr_1-\bfr_0, \bfr_2 - \bfr_0, \bfr_3-\bfr_0)(\bfR_1-\bfR_0, \bfR_2 - \bfR_0, \bfR_3-\bfR_0)^{-1}$. Given the constitutive law, one may then compute the energy density $W(\bflambda)$ for the tetrahedra, and hence the total energy of the LCE slab, $E$. Similarly, for each tetrahedron, one may compute the PK1 stress from $\mathbf{P}=\frac{\partial W}{\partial \bflambda}$, and the Cauchy stress follows as:
\begin{align}
\boldsymbol{\sigma}=\mathrm{Det}{(\bflambda)}^{-1} \mathbf{P} \bflambda^{\T}.
\end{align}
The elastic force on each node is then the gradient of the energy with respect to position: $\mathbf{f}_i=\partial E/\partial{\mathbf{R}_i}$. A simple calculation reveals that this derivative exactly corresponds to determining the surface traction at each face of the tetrahedron as $-\boldsymbol{\sigma} \bfn_i$, and then distributing these face forces equally amongst the three corresponding nodes. Thus, for each constitutive law, the code requires the stress function in order to simulate the slab.

Once the elastic forces are computed, the nodes are moved with standard damped Newtonian dynamics via symplectic Euler. Although the code is dynamical, the reported states are all static equilibria, except the coarsening in Fig.~4(a). For more details about the FE method, we refer the reader to the supplement of \cite{tallinen2013surface}.

\section{Creasing in a spontaneously elongating neo-Hooekan slab}

\begin{figure}[h]
	\includegraphics[width=\columnwidth]{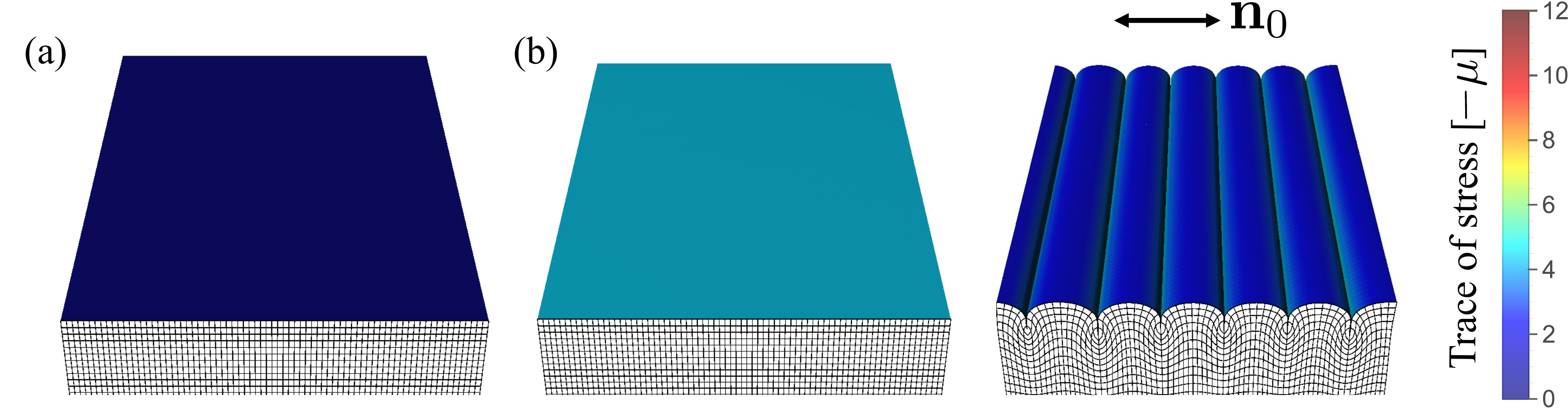}
	\caption{(a) Reference configuration with $\lambda_s=1$. (b) Equilibrium configurations under transversely isotropic compression $\lambda_s^{-1} \bfn_0 \otimes \bfn_0 + \sqrt{\lambda_s} (\bfI - \bfn_0 \otimes \bfn_0)$ with $\lambda_s=2$ and $\lambda_s = 2.42$ respectively.}
	\label{fig:comression}
\end{figure}

As introduced in the main text, the most naive model of the LCE surface instability treats the LCE as a standard neo-Hookean elastomer that spontaneously elongates by $\lambda_s$ along $\bfn_0=\hat{\mathbf{x}}$, leading to the energy density
\begin{equation}
W(\boldsymbol{\Lambda})=W_{NH}(\boldsymbol{\Lambda}\boldsymbol{\Lambda}_s^{-1}(\mathbf{x}))\mathrm{\ \ \ \ } \mathrm{Det}(\boldsymbol{\Lambda})=1 \label{eq:SI_neo}
\end{equation}
where $\boldsymbol{\Lambda}_s(\hat{\mathbf{x}}) = \lambda_s \hat{\mathbf{x}}\otimes\hat{\mathbf{x}}+\lambda_s^{-1/2} (\bfI-\hat{\mathbf{x}} \otimes \hat{\mathbf{x}} )$ is the spontaneous deformation gradient, and $W_{NH}(\boldsymbol{\Lambda})=\frac{1}{2} \mu \mathrm{Tr}(\boldsymbol{\Lambda} \boldsymbol{\Lambda}^\T)$ is the standard neo-Hookean function. Minimizing the total energy  is then  exactly equivalent to imposing a uniaxial mechanical compression $\boldsymbol{\Lambda}_s^{-1}(\mathbf{x})$ on a passive rubber slab, which is well known to produce the Biot creasing instability on the surface \cite{biot1965mechanics,hong2009formation,hohlfeld2011unfolding, tallinen2013surface}. In loading, following the (plane-strain) calculations of \cite{hong2009formation}, the point of linear-stability (Biot point) is at an equivalent applied compression of
\begin{equation}
\frac{\lambda_{zz}}{\lambda_{xx}}=3.4 \implies \frac{\sqrt{\lambda_s}}{1/\lambda_s}=3.4 \implies \lambda_s=2.26
\end{equation}
at which point cusped furrows emerge strongly sub-critically on the surface. In unloading, these cusped furrows remain the energy minimizer until the $T$ point at 
\begin{equation}
\frac{\lambda_{zz}}{\lambda_{xx}}=2.4 \implies \frac{\sqrt{\lambda_s}}{1/\lambda_s}=2.4 \implies \lambda_s=1.8
\end{equation}
The calculations in \cite{biot1965mechanics,hong2009formation,hohlfeld2011unfolding} all assume plane-strain (no $y$ displacement), guaranteeing the cusped creases form in lines perpendicular to $\bfn_0$. We therefore confirm these conclusions by conducting full 3D finite element simulations. The simulations are conducted on a slab with dimensions $L_x \times L_y \times L_z = 0.5 \times 0.5 \times 0.1$, discretized by the number of elements $(n_x,n_y,n_z)=(151,151,31)$. We  
substitute the energy (\ref{eq:SI_neo}) into the compressible treatment (\ref{eq:SI_compressible}) with $B=100\mu$ where $\mu$ is the shear modulus, compute the corresponding Cauchy stress analytically, and then apply the periodic boundary condition along $x$ and $y$, clamped boundary condition at $z=-t$ and free boundary condition at $z=0$. As shown in Fig.~\ref{fig:comression}, in the loading process, no surface instability is triggered at $\lambda_s=2$, but we see the stripe pattern when $\lambda_s$ exceeds the critical value. 

These considerations confirm that a uniaxially elongating rubber on a clamped foundation undergoes a classic Biot surface instability, producing cusped furrows that form, sub-critically, in parallel lines perpendicular to the director. This Biot instability lacks the key features of the LCE experiment --- cross-hatch pattern, oblique wave vector, smooth topography, supercritical --- and has a  threshold elongation $\lambda_s=2.26$ which exceeds that used experimentally $\lambda_s=2$. We thus turn to more sophisticated LCE constitutive laws to explain the observed instability.


\section{Soft Elasticity --- energies and stresses for the ideal and non-ideal LCEs}
The key additional physics for a realistic LCE constitutive law, is that the director can rotate within the LCE, generating soft elastic modes of deformation. Upon cooling, an LCE slab in the isotropic state can transform to the nematic state via a spontaneous deformation gradient $\bflambda_s(\bfn) = \lambda_s \bfn \otimes \bfn + \lambda_s^{-1/2}(\bfI - \bfn \otimes \bfn)$. If the subsequent deformation follows the neo-Hookean constitutive law,
the ideal LCE energy may be given by $W_{NH} \left( \bflambda \bflambda_s^{-1}(\bfn)\right)$,
where $\bflambda$ is the deformation gradient from the isotropic state to the nematic state and $W_{NH}$ is the neo-Hookean energy density given by $W_{NH}(\bflambda) = \frac{1}{2} \mu\Tr(\bflambda\bflambda^\T) $. This energy is identical to the previous section. However, now, additionally, we  allow the director $\bfn$ to rotate arbitrarily without any energy cost. Minimizing over $\bfn$ we thus obtain an ideal energy solely depending on $\bflambda$:
\beq
W_I(\bflambda) = \min_{\bfn} W_{NH} \left(\bflambda \bflambda_s^{-1} \right) = \frac{1}{2}\left(\Lambda_1^2 \lambda_s +  \Lambda_2^2 \lambda_2+ \Lambda_3^2/\lambda_s^2 \right), \mathrm{\ \ \ \ \ \ \ }\mathrm{Det}(\bflambda)=\Lambda_1\Lambda_2\Lambda_3=1
\eeq
where $\Lambda_1\leq \Lambda_2 \leq \Lambda_3$ are ordered principal stretches of $\bflambda$. However, owing to the fabrication, the rotation of $\bfn$ may cost a small amount of energy characterized by a small non-ideality factor $\alpha$, leading to the non-ideal energy:
\begin{align}
    W_{NI}(\bflambda) =  \min_{\bfn}  \left( W_{NH}(\bflambda \bflambda_s^{-1}(\bfn)) + \frac{1}{2}\alpha \lambda_s \mu \mathrm{Tr}\left(\bflambda(I-\bfn_0 \otimes \bfn_0) \bflambda^T \bfn\otimes\bfn \right)\right).
\end{align}
To minimize over $\bfn$, we collect all terms containing $\bfn$ and rewrite $W_{NI}(\bflambda)$ as 
\beq
W_{NI}(\boldsymbol{\Lambda})=\min_{\bfn} \left(\frac{1}{2}\lambda_s \mu \Tr \left( \boldsymbol{\Lambda} \boldsymbol{\Lambda}^{\T} + \bfM\bfn \otimes \bfn  \right) \right),\label{eq:nienergy}
\eeq
where $\bfM =(\lambda_s^{-3} - 1) \boldsymbol{\Lambda} \boldsymbol{\Lambda}^{\T} + \alpha \boldsymbol{\Lambda}(\bfI - \bfn_0 \otimes \bfn_0) \boldsymbol{\Lambda}^{\T}$.
In the admissible domain of $\lambda_s\sim 1.2-2.5$ and $\alpha\sim 0.05-0.3$, $\bfM$ is negative definite, and thus the minimizer $\bfn$ lies along the eigenvector corresponding to the smallest negative eigenvalue of $\bfM$. Specifically, let the corresponding eigenvalues and eigenvectors of $\bfM$ be $m_i$ and $\bfu_i$ for $i=1,2,3$, and $m_1 \leq m_2 \leq m_3 <0$. Then $W_{NI}(\boldsymbol{\Lambda})$, after minimizing over $\bfn$ for a fixed $\bfn_0$, may be written as
\beq
W_{NI}(\boldsymbol{\Lambda}) = 
\frac{1}{2}\lambda_s \mu \left( \Tr (\boldsymbol{\Lambda} \boldsymbol{\Lambda}^{\T}) + m_1 \right).
\label{eq:SI_WNI}
\eeq

Next, we compute the first Piola–Kirchhoff (PK1) stress $\bfP_{NI}$ by differentiating the non-ideal energy $W_{NI}(\bflambda J^{-1/3})$ directly in the slightly compressible regime ($J=\mathrm{Det}\bflambda \approx 1$). Notice that $m_1$ is also a function of $\bflambda$. Thus the PK1 stress includes the contribution of $m_1$ and is given by
\beq
\bfP_{NI}(\bflambda) = \frac{\partial W_{NI}(\boldsymbol{\Lambda}J^{-1/3})}{\partial \boldsymbol{\Lambda}} = \mu \lambda_s J^{-2/3} \left(\boldsymbol{\Lambda}  + m_1 \bfu_1 \otimes \boldsymbol{\Lambda}^{-1} \bfu_1 -\frac{1}{3} \left(\Tr(\boldsymbol{\Lambda} \boldsymbol{\Lambda}^{\T}) + m_1 \right)\boldsymbol{\Lambda}^{-\T} \right). \label{eq:PK_NI}
\eeq
Accordingly, the Cauchy stress $\boldsymbol{\sigma}_{NI}(\bflambda)$ is expressed as
\beq
\boldsymbol{\sigma}_{NI}(\bflambda) = J^{-1}\bfP_{NI}(\bflambda) \boldsymbol{\Lambda}^{\T} = \mu \lambda_s J^{-5/3}  \left(\boldsymbol{\Lambda} \boldsymbol{\Lambda}^{\T} + m_1 \bfu_1 \otimes \bfu_1 -\frac{1}{3} \left(\Tr(\boldsymbol{\Lambda} \boldsymbol{\Lambda}^{\T}) + m_1 \right)\bfI \right).
\label{eq:SI_Cauchy}
\eeq

{\bf Remark 1: symmetry}. The non-ideal energy $W_{NI}(\bflambda)$ is transversely isotropic, i.e., $W_{NI}(\bflambda) = W_{NI}(\bfR \bflambda \bfR_{\bfn_0})$ with $\bfR$ being an arbitrary rotation and $\bfR_{\bfn_0}$ a rotation about $\bfn_0$. In contrast, the ideal energy is completely isotropic, $W_{NI}(\bflambda) = W_{NI}(\bfR_1 \bflambda \bfR_{2})$, since no director is preferred.

{\bf Remark 2: Relations to energies in other works.} The ideal energy $W_{NI}(\bflambda)$ in this paper is identical to the well-known ideal ``trace formula" with $\lambda_s = r^{1/3}$.The latter is directly derived from statistical mechanics \cite{warner2007liquid}, assuming Gaussian polymer chains and uniform coupling between the chains and the nematic phase. With the same identification, our non-ideal form also exactly matches the well-established non-ideal trace formula \cite{warner2007liquid}, which was also originally derived statistically by additionally assuming cross-linking occurs in the nematic state with ``compositional fluctuations'' between chains in the strength of their coupling to the nematic field. The microstructural relaxation of this non-ideal energy was previously studied for thin sheets in tension in \cite{conti2002semisoft}. In that paper, the authors introduced a more compact form for the energy by writing it in terms of deformations $ \tilde{\bflambda}$ that are with respect to a different reference state. Precisely, $ \tilde{\bflambda}$ is related to our deformation gradient by $\bflambda = \tilde{\bflambda} b^{-1/6} (\bfI + (b^{1/2}-1) \bfn_0\otimes \bfn_0)$ where $b=1+\alpha \lambda_s^3/(1-\lambda_s^3)$. Substituting this into the standard non-ideal energy allows it to be written as:
\beq
W_{NI}(\tilde{\bflambda}) = \min_{\bfn}\frac{1}{2}\mu\left(\Tr(\tilde{\bflambda}\tilde{\bflambda}^\T) - \tilde{\alpha} |\tilde{\bflambda}^\T \bfn|^2 - \tilde{\beta} |\tilde{\bflambda}\bfn_0|^2\right),
\eeq
where $\tilde{\alpha} = 1-\alpha - \lambda_s^{-3}$ and $\tilde{\beta} = - \alpha \lambda_s^3/(1-\lambda_s^3)$.

\section{Finite elements with the non-ideal LCE energy}
\begin{figure}[h]
	\includegraphics[width=\columnwidth]{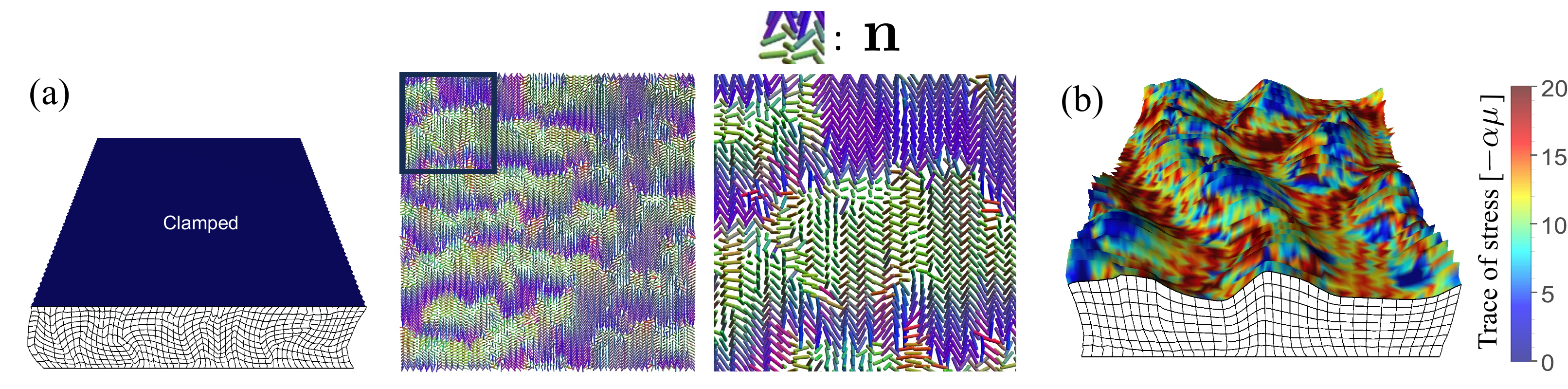}
	\caption{(a) Simulation of cooling an LCE slab from its isotropic state to nematic state with clamped top and bottom boundaries using $W_{NI}$ and distribution of directors at the middle surface. The color scheme represents the director's orientation. We see the mesh-scale oscillation of directors and the double laminates. (b) Corresponding simulation with free top surface and clamped bottom surface. The color scheme represents the trace of compressive stress at the top surface.}
	\label{fig:director}
\end{figure} 
We next consider simulating the instability directly using the non-ideal energy $W_{NI}$. The energy and the PK1 stress can be given analytically by Eq.~\ref{eq:SI_WNI} and Eq.~\ref{eq:SI_Cauchy} after minimizing over $\bfn$. It is thus straightforward to programme this stress into our FEM code (\ref{sec:fem}) to compute the  the equilibrium configuration as well as the final-state director distribution. We simulate a slab with dimensions $L_x \times L_y \times L_z = 0.5 \times 0.5 \times 0.1$, discretized by the number of elements $(n_x,n_y,n_z)=(101,101,21)$. Again the compressible energy (\ref{eq:SI_compressible}) is employed with $B=100\mu$, replacing $W(\bflambda)$ with $W_{NI}(\bflambda)$.

For an initial calculation, we  clamp both the top and bottom surfaces and then cool the LCE slab from its isotropic state to the nematic state, by slowly increasing $\lambda_s$ from $\lambda_s=1$ (isotropic) to $\lambda_s=2$ (final state). As shown in Fig.~\ref{fig:director}(a), this results in microstructure in the simulated LCE, with directors and deformations oscillating at the mesh-scale. The color scheme, which represents the directors' orientation, suggests  that double laminates emerge, consistent with the experimental observation in Fig.~\ref{fig:microstructure}, and our theoretical construction.

Relaxing the top surface produces thickness-scale topography that, charitably, illustrates the feature of the oblique wave vector in the surface instability (Fig.~\ref{fig:director}(b)). However, due to the alternating directors at the mesh scale, the deformation gradients and the surface stress oscillate rapidly across the domain, as indicated by the color scheme. 
Experimental evidence confirms that microstructures are formed at a length scale much smaller than the topography or the computationally accessible mesh size (green frame in Fig.~\ref{fig:microstructure}). Consequently, we turn to the two-scale computational approach discussed in the main text.


\section{Relaxation of the non-ideal LCE energy}

By forming compatible laminates, the ideal and non-ideal energies are convexified to relaxed energies. As a result, the LCEs exhibit soft and semi-soft mechanical responses. The relaxed ideal energy in 3D and the relaxed non-ideal energy in 2D  have been studied \cite{desimone2002macroscopic,conti2002semisoft}. However, the study of relaxed non-ideal energy in 3D is largely absent, hindering the understanding of 3D mechanical behavior, including surface instability in this work.

\begin{figure}[ht]
	\includegraphics[width=\columnwidth]{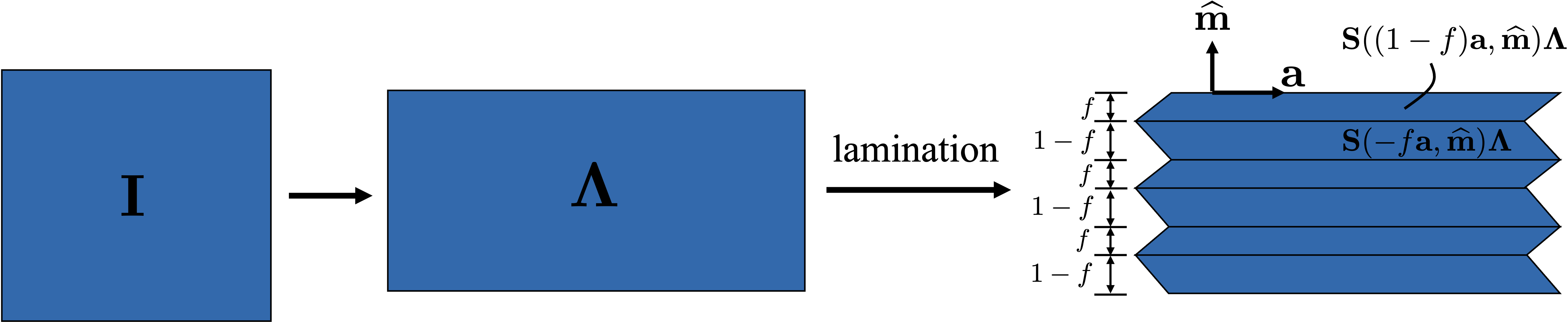}
	\caption{Approximating the deformation gradient $\bflambda$ with first-order twin.}
	\label{fig:lamination}
\end{figure}

Here we compute the relaxed non-ideal energy in 3D numerically and use it to construct a new constitutive relation for the study of surface instability. To this end, we firstly compute the relaxed energy $W_{R1}(\bflambda)$ by forming first-order laminates. A general shear is defined as $\bfS(\boldsymbol{\gamma}, \hat{\bfm}) = \bfI + \boldsymbol{\gamma}\otimes \hat{\bfm}$ with $\hat{\bfm}$ being the normal to the shear plane. $\boldsymbol{\gamma}$ is perpendicular to $\hat{\bfm}$ and encodes both the shear direction and amplitude. 
As shown in Fig.~\ref{fig:lamination}, a deformation gradient $\bflambda$ may be approximated by averaging two rank-1 deformation gradients $\bflambda_1$ and $\bflambda_2$ as
\beq
\bflambda = f \bflambda_1 + (1-f) \bflambda_2,    
\eeq
where $\bflambda_1 = \bfS((1-f)\bfa,\hat{\bfm}) \bflambda$ and $\bflambda_2 = \bfS(-f\bfa, \hat{\bfm}) \bflambda$. 
Then the relaxed energy $W_{R1}(\bflambda)$ by minimizing over laminates is given by
\beq
W_{R1}(\boldsymbol{\Lambda}) = \min_{f, \hat{\bfm}, \bfa \perp \hat{\bfm}} \left(f W_{NI}(\bflambda_1) + (1-f) W_{NI}(\bflambda_2) \right). \label{eq:R1minimization}
\eeq
Following the same procedure, we construct $W_{R2}(\bflambda)$ by forming second-order twins:
\beq
W_{R2}(\boldsymbol{\Lambda}) = \min_{f, \hat{\bfm}, \bfa \perp \hat{\bfm}} \left(f W_{R1}(\bflambda_1) + (1-f) W_{R1}(\bflambda_2) \right), \label{eq:R2minimization}
\eeq
replacing $W_{NI}(\bflambda)$ in (\ref{eq:R1minimization}) by $W_{R1}(\bflambda)$.
Practically in the numerical minimization, we write $\hat{\bfm}$ in terms of its Euler angles as
\beq
\hat{\bfm} = (\cos n_\theta, \sin n_\theta \cos n_\phi, \sin n_\theta \sin n\phi),
\eeq
and $\bfa = s \hat{\bfa}$ with magnitude $s$ and direction
\beq
\hat{\bfa} = \cos m_\phi (0, -\sin n_\phi, \cos n_\phi) + \sin m_\phi (-\sin n_\phi, \cos n_\theta \cos n_\phi, \cos n_\theta \sin n_\phi).
\eeq
Then we may perform the minimizations (\ref{eq:R1minimization}) and (\ref{eq:R2minimization}) over the generalized vector $ \boldsymbol{\ell} = (f, n_\theta, n_\phi, m_\phi, s)$ in the domain $\mathcal{D} =[0,1] \times [0, \pi] \times [0, 2\pi] \times [0, 2\pi] \times \mathbb{R}$. Here the amplitude of shear $s$ is restricted in a finite range for a sensible result, making $\mathcal{D}$ a finite domain.

To discretize the domain of $\bflambda$, we notice that $W_{R1}(\bflambda)$ is invariant under any rotation about $\bfn_0$. Obeying incompressibility and frame indifference, we may represent $\bflambda$ by any four independend scalar invariants. 
 Herein, we use $(\Lambda_1,\Lambda_3,\theta,\phi)$, with
$\Lambda_1 \leq \lambda_2 \leq \Lambda_3$ being the three eigenvalues of $\sqrt{\boldsymbol{\Lambda}^{\T}\boldsymbol{\Lambda}}$, $\theta$ the angle between $\bfn_0$ and the plane spanned by corresponding eigenvectors $\bfu_1$ and $\bfu_3$,  and $\phi$ the angle between $\bfu_1$ and the projected vector of $\bfn_0$ on that plane. If $\theta = \pi/2$, we define $\phi=0$.  Owing to the $\bfu_i \rightarrow -\bfu_i$ invariance, we may restrict $\theta$ and $\phi$ in $[0, \pi/2]$. Finally, we discretize the 4D space $(\Lambda_1, \Lambda_3, \theta, \phi) \in [0,1]\times[1,4]\times[0,\pi/2]\times[0,\pi/2]$ using $11 \times 31 \times 11 \times 11 = 41261$ nodes. 
For each node, we construct a test deformation gradient $\bflambda_t (\Lambda_1,\Lambda_3, \theta, \phi)$ and then perform the minimization for $\bflambda_t$.

The detailed numerical method is as follows. 
At each grid point $(\Lambda_1, \Lambda_3, \theta, \phi)$, we construct a test deformation gradient by 
\beqs
\boldsymbol{\Lambda}_t(\Lambda_1,\Lambda_3, \theta, \phi) &=& \Lambda_1 \bfu_1 \otimes \bfu_1 + \frac{1}{\Lambda_1 \Lambda_3} \bfu_2 \otimes \bfu_2 + \Lambda_3 \bfu_3 \otimes \bfu_3, \\
\text{where}~\bfu_1 &=& (\cos\theta \cos\phi, -\sin\phi, \sin\theta \cos\phi), \nonumber \\ 
\bfu_2 &=& (-\sin\theta, 0, \cos\theta), \nonumber \\ 
\bfu_3 &=& (-\cos\theta \sin\phi, -\cos\phi, -\sin\theta \sin\phi). \nonumber
\eeqs
 We then perform the minimization \ref{eq:R1minimization} numerically  in Mathematica, using a sequence of steps to ensure accurate answers throughout the domain, including areas where the homogeneous deformation is itself a local minimum.
 \begin{enumerate}
\item  Minimize at each grid point using NMimimize with the SimulatedAnnealing method,  a global minimization in the finite 5D domain $\mathcal{D}$, returning a minimized energy $W_{R1}(\boldsymbol{\Lambda})$ and its corresponding argument $ \boldsymbol{\ell} = (f, n_\theta, n_\phi, m_\phi, s)$, which is then saved as the current best microstructure at each grid point.

\item Minimize at each grid point using FindMinimum with ConjugateGradient method. This is a local minimization from a starting guess, and descends by evaluating the local gradient. At each grid point we take as initial guesses the result from step (1), and also 30 other starting points distributed over $\mathcal{D}$. The  saved microstrucure/energy at each point is updated  to the lowest of these results.

\item FindMinimum at each grid point, using as starting points the current current best microstructure parameters at the eight adjacent grid points. If the best result obtained is lower than the saved microstructure, it is updated. This step allows optimal microstructures to propagate across the grid, and is  repeated until convergence is achieved over the entire grid.
\end{enumerate}
 
 We find that the ConjugateGradient steps give very strong convergence to local minima if we provide  analytic forms for the  gradient $\frac{\partial W_{R1}(\boldsymbol{\Lambda}_t)}{\partial \boldsymbol{\ell}}$ which, for completeness, are displayed below:
\beqs
\frac{\partial W_{R1}(\boldsymbol{\Lambda}_t)}{\partial f} &=& W_{NI}(\bflambda_1) - W_{NI}(\bflambda_2) - f \bfa \cdot \boldsymbol{\sigma}_{NI}(\bflambda_1)\mhat - (1-f) \bfa \cdot \boldsymbol{\sigma}_{NI}(\bflambda_2)\mhat \nonumber \\
\frac{\partial W_{R1}(\boldsymbol{\Lambda}_t)}{\partial n_{\theta}} &=& f(1-f)\bfa\cdot \left( \bfP_{NI}(\bflambda_1) -  \bfP_{NI}(\bflambda_2)\right)\bflambda_t^\T \frac{\partial \mhat}{\partial n_{\theta}} + f(1-f)\left(\boldsymbol{\sigma}_{NI}(\bflambda_1) - \boldsymbol{\sigma}_{NI}(\bflambda_2)\right)\mhat \cdot \frac{\partial\bfa}{\partial n_{\theta}} \nonumber \\
\frac{\partial W_{R1}(\boldsymbol{\Lambda}_t)}{\partial n_{\phi}} &=& f(1-f)\bfa\cdot \left( \bfP_{NI}(\bflambda_1) -  \bfP_{NI}(\bflambda_2)\right)\bflambda_t^\T \frac{\partial \mhat}{\partial n_{\phi}} + f(1-f)\left(\boldsymbol{\sigma}_{NI}(\bflambda_1) - \boldsymbol{\sigma}_{NI}(\bflambda_2)\right)\mhat \cdot \frac{\partial\bfa}{\partial n_{\phi}} \nonumber \\
\frac{\partial W_{R1}(\boldsymbol{\Lambda}_t)}{\partial m_{\phi}} &=& f(1-f)(\boldsymbol{\sigma}_{NI}(\bflambda_1)-\boldsymbol{\sigma}_{NI}(\bflambda_2))\mhat \cdot \frac{\partial \bfa}{\partial m_{\phi}} \nonumber \\
\frac{\partial W_{R1}(\boldsymbol{\Lambda}_t)}{\partial s} &=& f(1-f)(\boldsymbol{\sigma}_{NI}(\bflambda_1)-\boldsymbol{\sigma}_{NI}(\bflambda_2))\mhat \cdot \ahat 
\eeqs

The resulting grid of energies $W_{R1}(\Lambda_1, \Lambda_3, \theta, \phi)$ and microstructures  $ \boldsymbol{\ell}_1(\Lambda_1, \Lambda_3, \theta, \phi)$ are interpolated at cubic order in Mathematica, resulting in interpolated functions $f_{W_{R1}}(\Lambda_1, \Lambda_3, \theta, \phi)$ and $f_{\boldsymbol{\ell}_1}(\Lambda_1, \Lambda_3, \theta, \phi)$. A numerical approximation for any deformation gradient $\tilde{\boldsymbol{\Lambda}}$, is then given by first calculating the four parameters $(\tilde{\Lambda}_1,\tilde{\Lambda}_3,\tilde{\theta},\tilde{\phi})$ for $\tilde{\boldsymbol{\Lambda}}$ and then evaluating the function $f_{W_{R1}}$ to obtain its energy $W_{R1} (\tilde{\boldsymbol{\Lambda}}) = f_{W_{R1}}(\tilde{\Lambda}_1,\tilde{\Lambda}_3,\tilde{\theta},\tilde{\phi})$, and similar for microstructure parameters $\boldsymbol{\ell}_1(\tilde{\boldsymbol{\Lambda}})$.

 The computation of  $W_{R2}(\bflambda_t)$ then follows the same procedure as above, except replacing $W_{NI}$ by the numerical  $W_{R1}(\boldsymbol{\Lambda})$, and using numerical derivatives of $W_{R1}(\boldsymbol{\Lambda})$ during ConjugateGradient. The result is again a grid of  energies $W_{R2}(\Lambda_1, \Lambda_3, \theta, \phi)$ and microstructures  $ \boldsymbol{\ell}_2(\Lambda_1, \Lambda_3, \theta, \phi)$.

The lack of analytic derivatives, combined with numerical inaccuracies in $W_{R1}(\boldsymbol{\Lambda})$, mean that the R2 microstructures thus obtained are often unsatisfactorily converged into local minima, particularly for the subsequent estimation of stresses which relies on strong local convergence (see next section). We thus then perform, at each grid point, an additional ConjugateGradient minimization (\ref{eq:W2_full}) for $W_{R2}$ as the sum of four monodomain energies, with minimization occurring simultaneously over the 15 parameters of a full second-order twin:
\beq
W_{R2}(\boldsymbol{\Lambda}) = \min_{\boldsymbol{\ell_2},\boldsymbol{\ell_{11}},\boldsymbol{\ell_{12}} \in \mathcal{D}} f_2 f_{11} W_{NI}(\bflambda_{11})+  f_2 (1-f_{11}) W_{NI}(\bflambda_{12})
    +
    (1-f_2) f_{12} W_{NI}(\bflambda_{21}) + (1-f_2) (1-f_{12}) W_{NI}(\bflambda_{22}) \label{eq:W2_full}
\eeq
  This final minimization starts from an initial value of $(\boldsymbol{\ell}_2, \boldsymbol{\ell}_{11},\boldsymbol{\ell}_{12}) =(\boldsymbol{\ell}_2(\boldsymbol{\Lambda}_t), \boldsymbol{\ell}_{1}(\boldsymbol{\Lambda}_1),\boldsymbol{\ell}_{1}(\boldsymbol{\Lambda}_2))$, where $\boldsymbol{\ell}_2$ comes from the grid of micro-structures obtained in the previous $W_{R2}$ calculation, while $\boldsymbol{\ell}_{1}(\boldsymbol{\Lambda}_1)$ and $\boldsymbol{\ell}_{1}(\boldsymbol{\Lambda}_2)$ are the R1 microstructure interpolation functions evaluated at the deformation gradients in the construction $\boldsymbol{\ell}_2(\boldsymbol{\Lambda}_t)$. In this case, we may indeed derive the relevant gradients analytically to obtain strong local convergence:
\begin{align}
        \frac{\partial W_{R2}(\boldsymbol{\Lambda})}{\partial f_2} &= f_{11} W_{NI}(\bflambda_{11}) + (1-f_{11})W_{NI}(\bflambda_{12}) - f_{12} W_{NI}(\bflambda_{21}) - (1-f_{12}) W_{NI}(\bflambda_{22}) \nonumber \\ &- \frac{\bfa_2\cdot dW_1dF_2 \cdot \mhat_2}{1-f2} - \frac{\bfa_2 \cdot dW_2dF_2\cdot \mhat_2}{f_2} \nonumber \\
    \frac{\partial W_{R2}(\boldsymbol{\Lambda})}{\partial n_{\theta2}}&=  \bfa_2\cdot (dW_1dF_2 - dW_2dF_2 ) \cdot \left.\frac{\partial \mhat}{\partial \theta}\right|_{\boldsymbol{\ell}_2} + (dW_1dF_2 - dW_2dF_2) \cdot \mhat_2 \cdot \left.\frac{\partial \bfa}{\partial n_{\theta}}\right|_{\boldsymbol{\ell}_2} \nonumber  \\
    \frac{\partial W_{R2}(\boldsymbol{\Lambda})}{\partial n_{\phi2}}&= \bfa_2\cdot (dW_1dF_2 -  dW_2dF_2 ) \cdot \left.\frac{\partial \mhat}{\partial n_\phi}\right|_{\boldsymbol{\ell}_2} + (dW_1dF_2  - dW_2dF_2 )\cdot \mhat_2 \cdot \left.\frac{\partial \bfa}{\partial n_\phi}\right|_{\boldsymbol{\ell}_2}            \nonumber\\
    \frac{\partial W_{R2}(\boldsymbol{\Lambda})}{\partial m_{\phi 2}}&=  (dW_1dF_2  - dW_2dF_2 )\cdot \mhat_2 \cdot \left.\frac{\partial \bfa}{\partial m_\phi}\right|_{\boldsymbol{\ell}_2} \nonumber \\
    \frac{\partial W_{R2}(\boldsymbol{\Lambda})}{\partial s_2}&= (dW_1dF_2  - dW_2dF_2) \cdot \mhat_2 \cdot \left.\frac{\partial \bfa}{\partial s}\right|_{\boldsymbol{\ell}_2},\nonumber \\
        \frac{\partial W_{R2}(\boldsymbol{\Lambda})}{\partial f_{11}} &= f_{2} W_{NI}(\bflambda_{11})  - f_{2} W_{NI}(\bflambda_{12}) - \frac{\bfa_{11}\cdot dW_{11}dF_{11}\cdot \mhat_{11}}{1-f_{11}} - \frac{\bfa_{11}\cdot dW_{12}dF_{11}\cdot\mhat_{11}}{f_{11}} \nonumber \\
    \frac{\partial W_{R2}(\boldsymbol{\Lambda})}{\partial n_{\theta 11}}&=  \bfa_{11}\cdot( dW_{11}dF_{11} - dW_{12}dF_{11} ) \cdot \left.\frac{\partial \mhat}{\partial \theta}\right|_{\boldsymbol{\ell}_{11}} + (dW_{11}dF_{11} - dW_{12}dF_{11}) \cdot \mhat_{11} \cdot \left.\frac{\partial \bfa}{\partial n_{\theta}}\right|_{\boldsymbol{\ell}_{11}} \nonumber  \\
    \frac{\partial W_{R2}(\boldsymbol{\Lambda})}{\partial n_{\phi 11}}&= \bfa_{11}\cdot( dW_{11}dF_{11} - dW_{12}dF_{11} ) \cdot \left.\frac{\partial \mhat}{\partial n_\phi}\right|_{\boldsymbol{\ell}_{11}} + (dW_{11}dF_{11} - dW_{12}dF_{11}) \cdot \mhat_{11} \cdot \left.\frac{\partial \bfa}{\partial n_\phi}\right|_{\boldsymbol{\ell}_{11}}           \nonumber\\
    \frac{\partial W_{R2}(\boldsymbol{\Lambda})}{\partial m_{\phi 11}}&=  (dW_{11}dF_{11}  - dW_{12}dF_{11}) \cdot \mhat_{11} \cdot \left.\frac{\partial \bfa}{\partial m_\phi}\right|_{\boldsymbol{\ell}_{11}} \nonumber \\
    \frac{\partial W_{R2}(\boldsymbol{\Lambda})}{\partial s_{11}}&= (dW_{11}dF_{11}  - dW_{12}dF_{11}) \cdot \mhat_{11} \cdot \left.\frac{\partial \bfa}{\partial s}\right|_{\boldsymbol{\ell}_{11}}, \nonumber \\
        \frac{\partial W_{R2}(\boldsymbol{\Lambda})}{\partial f_{12}} &= (1-f_{2}) W_{NI}(\bflambda_{21})  - (1-f_{2}) W_{NI}(\bflambda_{22}) - \frac{\bfa_{12}\cdot dW_{21}dF_{11}\cdot \mhat_{12}}{1-f_{12}} - \frac{\bfa_{12}\cdot dW_{22}dF_{11}\cdot\mhat_{12}}{f_{12}} \nonumber \\
    \frac{\partial W_{R2}(\boldsymbol{\Lambda})}{\partial n_{\theta 12}}&=  \bfa_{12}\cdot( dW_{21}dF_{11} - dW_{22}dF_{11} ) \cdot \left.\frac{\partial \mhat}{\partial \theta}\right|_{\boldsymbol{\ell}_{12}} + (dW_{21}dF_{11} - dW_{22}dF_{11}) \cdot \mhat_{11} \cdot \left.\frac{\partial \bfa}{\partial n_{\theta}}\right|_{\boldsymbol{\ell}_{12}} \nonumber  \\
    \frac{\partial W_{R2}(\boldsymbol{\Lambda})}{\partial n_{\phi 12}}&= \bfa_{12}\cdot( dW_{21}dF_{11} - dW_{22}dF_{11} ) \cdot \left.\frac{\partial \mhat}{\partial n_\phi}\right|_{\boldsymbol{\ell}_{12}} + (dW_{21}dF_{11} - dW_{22}dF_{11}) \cdot \mhat_{11} \cdot \left.\frac{\partial \bfa}{\partial n_\phi}\right|_{\boldsymbol{\ell}_{12}}          \nonumber\\
    \frac{\partial W_{R2}(\boldsymbol{\Lambda})}{\partial m_{\phi 12}}&=  (dW_{21}dF_{11} \cdot - dW_{22}dF_{11}) \cdot \mhat_{12}) \cdot \left.\frac{\partial \bfa}{\partial m_\phi}\right|_{\boldsymbol{\ell}_{12}} \nonumber \\
    \frac{\partial W_{R2}(\boldsymbol{\Lambda})}{\partial s_{12}}&= (dW_{21}dF_{11}  - dW_{22}dF_{11}) \cdot \mhat_{12}) \cdot \left.\frac{\partial \bfa}{\partial s}\right|_{\boldsymbol{\ell}_{12}}
\end{align}
 where the intermediate derivatives are
 \begin{align}
dW_1dF_2 &= f_2(1-f_2)\left(f_{11}\bfS(\mhat, (1-f)\bfa)|_{\boldsymbol{\ell}_{11}} \cdot \bfP_{NI}(\bflambda_{11})\cdot\bflambda^\T + (1-f_{11})\bfS(\mhat, -f\bfa)|_{\boldsymbol{\ell}_{11}} \cdot \bfP_{NI}(\bflambda_{12})\cdot\bflambda^\T\right)    \nonumber\\
dW_2dF_2 &=  f_2(1-f_2)\left(f_{12}\bfS(\mhat, (1-f)\bfa)|_{\boldsymbol{\ell}_{12}} \cdot \bfP_{NI}(\bflambda_{21})\cdot\bflambda^\T + (1-f_{12})\bfS(\mhat, -f\bfa)|_{\boldsymbol{\ell}_{12}} \cdot \bfP_{NI}(\bflambda_{22})\cdot\bflambda^\T\right)    \nonumber\\
dW_{11}dF_{11} &= f_2 f_{11} (1-f_{11})\left(\bfP_{NI}(\bflambda_{11})  \bflambda_1^\T\right)  \nonumber\\
dW_{12}dF_{11} &= f_2 f_{11} (1-f_{11})\left(\bfP_{NI}(\bflambda_{12})  \bflambda_1^\T\right)   \nonumber\\
dW_{21}dF_{11} &= (1-f_2) f_{12} (1-f_{12})\left(\bfP_{NI}(\bflambda_{21})  \bflambda_2^\T\right)   \nonumber\\
dW_{22}dF_{11} &=  (1-f_2) f_{12} (1-f_{12})\left(\bfP_{NI}(\bflambda_{21})  \bflambda_2^\T\right)  
 \end{align}
and the deformation gradients are 
 \beqs
&& \bflambda_1= \bfS((1-f)\bfa, \mhat)|_{\boldsymbol{\ell}_2} \bflambda,\quad \bflambda_2= \bfS(-f\bfa, \mhat)|_{\boldsymbol{\ell}_2} \bflambda,\nonumber\\
 &&\bflambda_{11} = \bfS((1-f)\bfa, \mhat)|_{\boldsymbol{\ell}_{11}} \bfS((1-f)\bfa, \mhat)|_{\boldsymbol{\ell}_{2}} \bflambda,\quad   \bflambda_{12} = \bfS(-f \bfa, \mhat)|_{\boldsymbol{\ell}_{12}} \bfS((1-f)\bfa, \mhat)|_{\boldsymbol{\ell}_{2}} \bflambda, \nonumber \\  &&\bflambda_{21} = \bfS((1-f)\bfa, \mhat)|_{\boldsymbol{\ell}_{21}} \bfS(-f\bfa, \mhat)|_{\boldsymbol{\ell}_{2}} \bflambda, \quad  \bflambda_{22} = \bfS(-f\bfa, \mhat)|_{\boldsymbol{\ell}_{21}} \bfS(-f\bfa, \mhat)|_{\boldsymbol{\ell}_{2}} \bflambda.
 \eeqs
Finally, we again interpolate the resulting grid of energies and microstructures in mathematica at cubic order, to give the interpolated functions $W_{R2}(\boldsymbol{\Lambda})$ and  $(\boldsymbol{\ell}_2(\boldsymbol{\Lambda}),\boldsymbol{\ell}_{11}(\boldsymbol{\Lambda}),\boldsymbol{\ell}_{12}(\boldsymbol{\Lambda}))$.

\begin{figure}[ht]
	\includegraphics[width=.8\columnwidth]{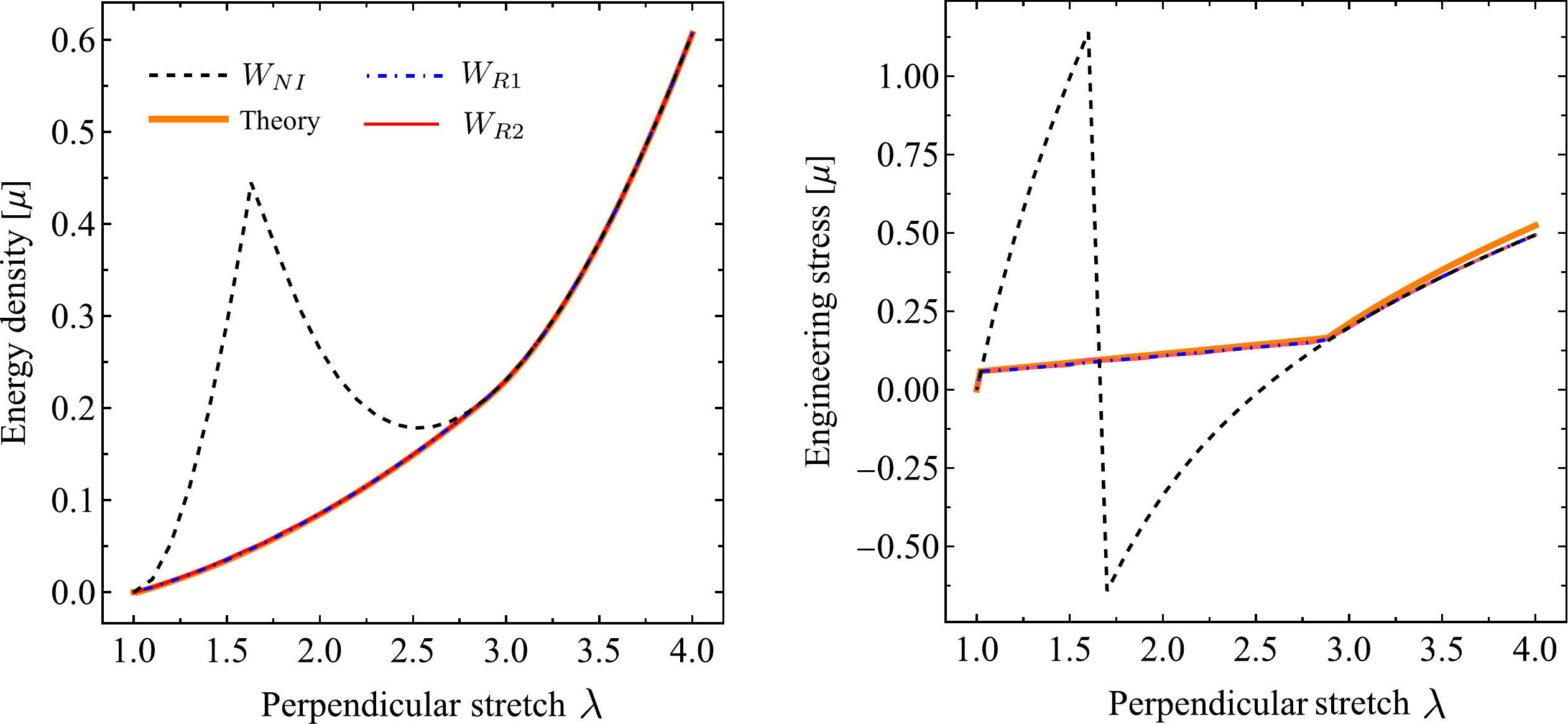}
	\caption{A benchmark test: stretching perpendicular to $\bfn_0$ for $(\lambda_s, \alpha)=(2,0.05)$.}
	\label{fig:stretching}
\end{figure}

 In addition to the two numerical experiments in the main text, we provide a further test of our results, in the well studied ``stripe-domain" geometry: an initially relaxed thin sheet with in-plane director, that is stretched perpendicular to $\bfn_0$. More precisely, we consider a planar LCE in the $x-y$ plane, with $\bfn_0=\mathbf{\hat{y}}$, so that the minimizing spontaneous distortion is $\Lambda_s(\mathbf{\hat{y}})$. We then consider imposing a further perpendicular stretch by $\lambda$ in the $\mathbf{\hat{x}}$ direction. Since the sample is thin, it will relax in the thickness direction by an unknown amount $\lambda_{zz}$, giving the (incompressible) deformation  $\bfF=\mathrm{diag}(\lambda,1/(\lambda_{zz} \lambda),\lambda_{zz})$.  Minimizing over relaxations then gives the relevant energy as
 \begin{align}
W(\lambda)= \min_{\lambda_{zz}}W_{R2}\left(\mathrm{diag}(\lambda,1/(\lambda_{zz} \lambda),\lambda_{zz})\cdot \Lambda_s(\mathbf{\hat{y}})\right).
 \end{align}
This additional relaxation is characteristic of thin sheets and allows a complete analytic solution \cite{conti2002semisoft}. However, here we conduct it as a final minimization step in Mathematica, using our previously computed numerical $W_{R2}$. As seen in Fig.\ \ref{fig:stretching},  this procedure gives energy stress  in excellent agreement with the well-established analytic answer \cite{warner2007liquid}: a stringent test of our numerical relaxation.

\section{Computation of the stress}
To perform finite element simulations, it is necessary to have a constitutive law that describes the relationship between stress and strain for the relaxed energy $W_{R2}$. However, obtaining the stress through direct numerical differentiation of $W_{R2}$ can be challenging due to its numerical nature. Instead, we use an alternative method to calculate the PK1 stress of $W_{R2}(\boldsymbol{\Lambda})$ based on the four monodomain variants that make up the second-order twin. The PK1 stress of each of the four variants can be obtained analytically using the equation $\bfP_{NI}$ as shown in (\ref{eq:PK_NI}). The PK1 stress of $W_{R2}$ is then obtained by taking the volume average of the PK1 stresses of the four variants. This approach enables us to obtain the stress-strain relationship for use in finite element simulations.

We first prove that the PK1 stress of $W_{R1}$ is the volume average of the PK1 stresses of its variants.

\begin{theorem} \label{lemma1}
Suppose 
\beq
W_{R1}(\boldsymbol{\Lambda}) = \min_{f\in[0,1], \mhat, \bfa\cdot\mhat=0} \left( f W_{NI}\left(\bflambda + (1-f)\bfa \otimes \hat{\bfb}\right) + (1-f) W_{NI} \left(\bflambda - f \bfa \otimes \hat{\bfb} \right)\right)
\label{eq:thm1}
\eeq
is the relaxed energy of the first-order twin. Then the PK1 stresses, computed by taking derivatives of the energy with respect to deformation gradients, satisfy $\bfP_{R1}(\boldsymbol{\Lambda}) = f \bfP_{NI}(\bflambda_1) + (1-f) \bfP_{NI}(\bflambda_2) $, equivalently,
\beq
\frac{\d W_{R1}(\boldsymbol{\Lambda})}{\d \boldsymbol{\Lambda}} = f \left.\frac{\partial W_{NI}(\bflambda)}{\partial \bflambda}\right|_{\bflambda=\bflambda_1} + (1-f) \left.\frac{\partial W_{NI}(\bflambda)}{\partial \bflambda}\right|_{\bflambda=\bflambda_2} \label{eq:PK1_1}
\eeq
where $\bflambda_1 = \bflambda + (1-f)\bfa\otimes \hat{\bfb}$ and $\bflambda_2 = \bflambda - f  \bfa \otimes \hat{\bfb}$ are shears.
\end{theorem}
\begin{proof}
    Notice that for a given $\boldsymbol{\Lambda}$, the minization in (\ref{eq:thm1}) will provide the optimal parameters $(f, \bfa, \hat{\bfb})$ as functions of $\boldsymbol{\Lambda}$. Also, we use the normal $\hat{\bfb}$ to the reference twinning plane for simplicity.
  To simplify the notation, let us define $W_{R1}(\boldsymbol{\Lambda})$ as $g(f(\boldsymbol{\Lambda}), \bfa(\boldsymbol{\Lambda}), \hat{\bfb}(\boldsymbol{\Lambda}), \boldsymbol{\Lambda})$, where $g$ represents the right-hand side of (\ref{eq:thm1}) after minimization. Taking the derivative of $W_{R1}(\bflambda)$ with respect to $\bflambda$ yields
  \beq
  \frac{\d W_{R1}(\bflambda)}{\d \bflambda} = \frac{\partial g}{ \partial f} \frac{\partial f}{ \partial \bflambda} + \frac{\partial g}{ \partial \bfa} \cdot \frac{\partial \bfa}{ \partial \bflambda} + + \frac{\partial g}{ \partial \hat{\bfb}} \cdot \frac{\partial \hat{\bfb}}{ \partial \bflambda} + + \frac{\partial g}{ \partial \bfa} \cdot \frac{\partial \bfa}{ \partial \bflambda} + \frac{\partial g}{\partial \bflambda}. \label{eq:partial}
  \eeq

  Since $f$, $\bfa$, and $\bhat$ are the minimizers of $g$, we have:
    \beqs
    \frac{\partial g}{\partial f}=0 &\Leftrightarrow& W_{NI}(\bflambda_1) - W_{NI}(\bflambda_2) + \left(f \frac{\partial W_{NI}}{\partial \bflambda_1}  + (1-f) \frac{\partial W_{NI}}{\partial \bflambda_2}\right) :(-\bfa \otimes \bhat)  =0 \label{eq:g1} \\
    \frac{\partial g}{\partial \bfa} = 0 &\Leftrightarrow&f(1-f) \left( \frac{\partial W_{NI}}{\partial \bflambda_1} - \frac{\partial W_{NI}}{\partial \bflambda_2} \right)\bhat=0 \label{eq:g2}\\
    \frac{\partial g}{\partial \mhat} = 0 &\Leftrightarrow & f(1-f) \bfa \left(  \frac{\partial W_{NI}}{\partial \bflambda_1}-  \frac{\partial W_{NI}}{\partial \bflambda_2}\right)=0 \label{eq:g3}
    \eeqs
    where $\frac{\partial W_{NI}}{\partial \bflambda_1} = \left.\frac{\partial W_{NI}(\bflambda)}{\partial \bflambda}\right|_{\bflambda=\bflambda_1}$ and $\frac{\partial W_{NI}}{\partial \bflambda_2} = \left.\frac{\partial W_{NI}(\bflambda)}{\partial \bflambda}\right|_{\bflambda=\bflambda_2}$. Physically, Eqs.~(\ref{eq:g2}) and (\ref{eq:g3}) denote the continuity of force across the twinning plane.   
    Substituting the zero derivatives into (\ref{eq:partial}), the PK1 stress of $W_{R1}(\boldsymbol{\Lambda})$ may be computed by
    \beq
    \frac{\d W_1(\boldsymbol{\Lambda})}{\d \boldsymbol{\Lambda}} = \frac{\partial g}{\partial \boldsymbol{\Lambda}} = f \frac{\partial W_{NI}}{\partial \bflambda_1} + (1-f) \frac{\partial W_{NI}}{\partial \bflambda_2}.
    \eeq
This completes the proof. 
\end{proof}


An immediate generalization of Eq.~\ref{eq:PK1_1} implies that the PK1 stress for $W_{R2}(\bflambda)$ may be obtained by averaging the PK1 stresses of the four variants:
\beq
\bfP_{R2} (\bflambda) = f_2f_{11} \bfP_{NI}(\bflambda_{11}) + f_2(1-f_{11})\bfP_{NI}(\bflambda_{12}) + (1-f_2)f_{12} \bfP_{NI}(\bflambda_{21}) + (1-f_2)(1-f_{12})\bfP_{NI}(\bflambda_{22}),
\eeq
where the four deformations $\bflambda_{11}$, $\bflambda_{12}$, $\bflambda_{21}$ and $\bflambda_{22}$ are obtained from $W_{R2}(\bflambda)$.
In the numerical relaxation, we calculate the PK1 stress solely for the test deformation gradient $\bflambda_t$. Subsequently, we use interpolation to generate a PK1 stress function $\bfP_{PK1}(\Lambda_1, \Lambda_3, \theta, \phi)$ that corresponds to the test deformation gradient. To obtain the PK1 stress for an arbitrary $\bflambda$ that matches the parameters of $\bflambda_t$, we identify two orthogonal transformations $\bfR_1$ and $\bfR_2$ that satisfy $\bflambda = \bfR_1 \bflambda_t \bfR_2$. Then, we calculate the PK1 stress for $\bflambda$ as $\bfP_{R_2}(\bflambda) =\bfR_1 \bfP_{PK1}(\Lambda_1, \Lambda_3, \theta, \phi) \bfR_2$. Accordingly, the Cauchy stress is given by $\boldsymbol{\sigma}_{R2} = \bfP_{R2}(\bflambda) \bflambda^{\T}$ in the incompressible regime.

\section{Finite element method with the numerically relaxed non-ideal LCE energy}
To use our numerically relaxed LCE energy in FEM, we take the energy as :
\beq
W= f W_{R2} (\boldsymbol{\Lambda} J^{-1/3}  ) + (1-f) \frac{\mu}{2}\left[ 
\Tr(\boldsymbol{\Lambda} \boldsymbol{\Lambda}^{\T} J^{-2/3}) -3 \right] + \frac{1}{2}B (J -1)^2.
\eeq
where  $B=100\mu$ makes the LCE almost incompressible, and $0<f<1$ changes the LCE consitutive law from neo-Hookean (i.e. isotropic, $f=0$) to pure LCE ($f=1$). The corseponding Cauchy stress in the finite elements also contains three components: the LCE stress $\boldsymbol{\sigma}_L$, the neo-Hookean stress $\boldsymbol{\sigma}_N$, and the bulk pressure $\boldsymbol{\sigma}_B$, given by:
\beqs
\boldsymbol{\sigma}_L &=&  \left[ \tilde{\bfP}_2 \boldsymbol{\Lambda}^{\T} - \frac{1}{3}\Tr(
\tilde{\bfP}_2\boldsymbol{\Lambda}^{\T}
)\bfI \right] J^{-4/3} \\
\boldsymbol{\sigma}_N &=& \mu \left[\boldsymbol{\Lambda} \boldsymbol{\Lambda}^{\T} - \frac{1}{3}\Tr(\boldsymbol{\Lambda} \boldsymbol{\Lambda}^{\T} ) \bfI \right] J^{-5/3} \\
\boldsymbol{\sigma}_B &=& B (J-1)\bfI
\eeqs
where $\tilde{\bfP}_2 = \bfP_{R2}(\bflambda J^{-1/3})$.
The dynamics coarsening simulation starts directly with $f=1$, while all others start from the isotropic state and  are ``cooled'' quasistatically to the nematic state by slowly increasing $f$ from zero to one.

\begin{figure}[h]
	\includegraphics[width=\columnwidth]{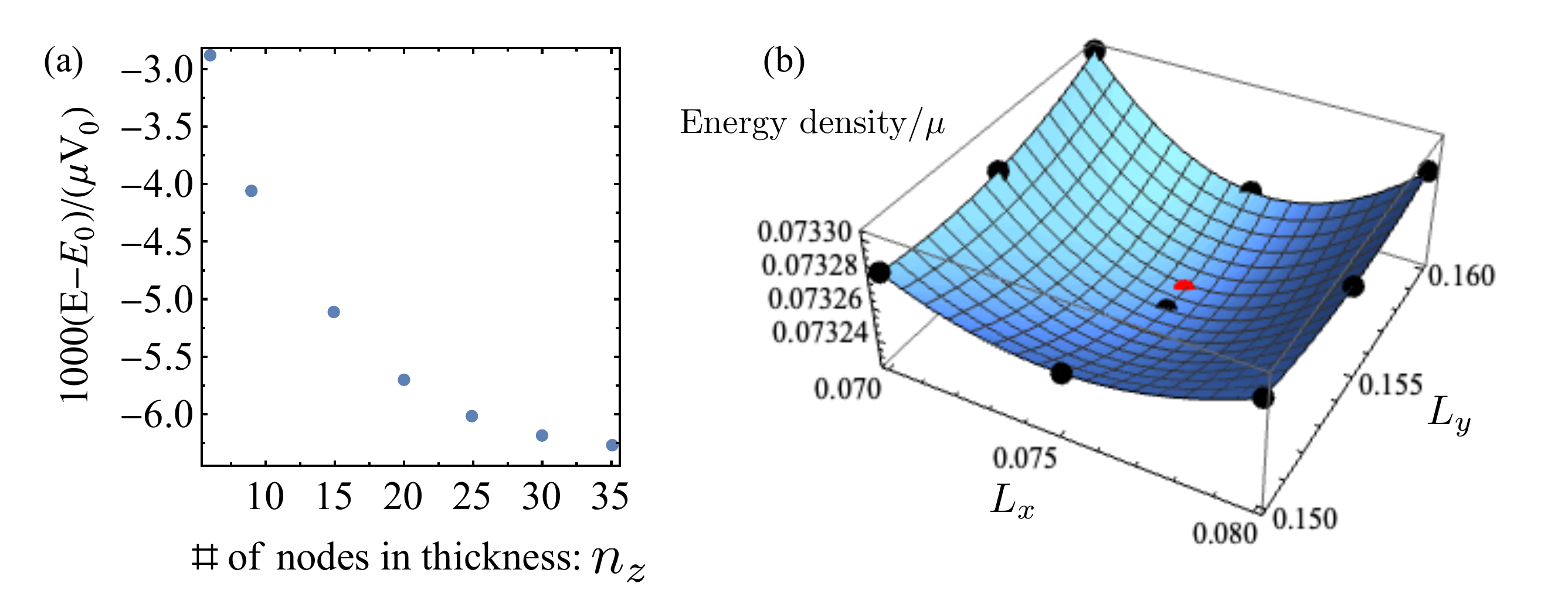}
	\caption{Mesh validation and energy interpolation.}
	\label{fig:validation}
\end{figure}

To ensure well converged results, we mapped the mesh size dependence and found that energy convergence is only  achieved when the number of nodes in the thickness direction  ($L_z$) exceeds 30 (Fig.~\ref{fig:validation}(a)). 

As seen in Fig 4, minimzing the $W_{R2}$ results in a 2D periodic pattern in the equilibrium state. To find the geometrically optimized unit-cell for the pattern,  we first simulated a large domain with dimensions $10t \times 10t \times t$ to roughly estimate the wavelengths $\tilde{a}_1$ and $\tilde{a}_2$ along $x$ and $y$. Next, we conducted nine calibrated simulations with a finer mesh. We set the simulated domains $L_x \times L_y \times t$ to be adjacent to $\tilde{a}_1\times \tilde{a}_2\times t$, specifically $L_x\times L_y \times t =(\tilde{a}_1-0.1 t, \tilde{a}_1, \tilde{a}_1 + 0.1t) \times (\tilde{a}_2 - 0.1t, \tilde{a}_2, \tilde{a}_2 + 0.1t) \times t $. We plotted the simulated energy densities (energy per unit reference volume) against $(L_x,L_y)$ in Fig.\ref{fig:validation}(b), which formed a convex envelope including an energy minimum. If the energy minimum point was not included, we adjusted $(\tilde{a}_1,\tilde{a}_2)$ and repeated the process. We then interpolated the energy density to determine the optimal wavelengths $(L_x,L_y)=(a_1,a_2)$ along $x$ and $y$ corresponding to the energy minimum. Finally, we simulated the equilibrium state of the domain size $(a_1,a_2,t)$ with a finer mesh and then periodically translated it to obtain a larger domain for illustration.

\begin{figure}[h]
	\includegraphics[width=\columnwidth]{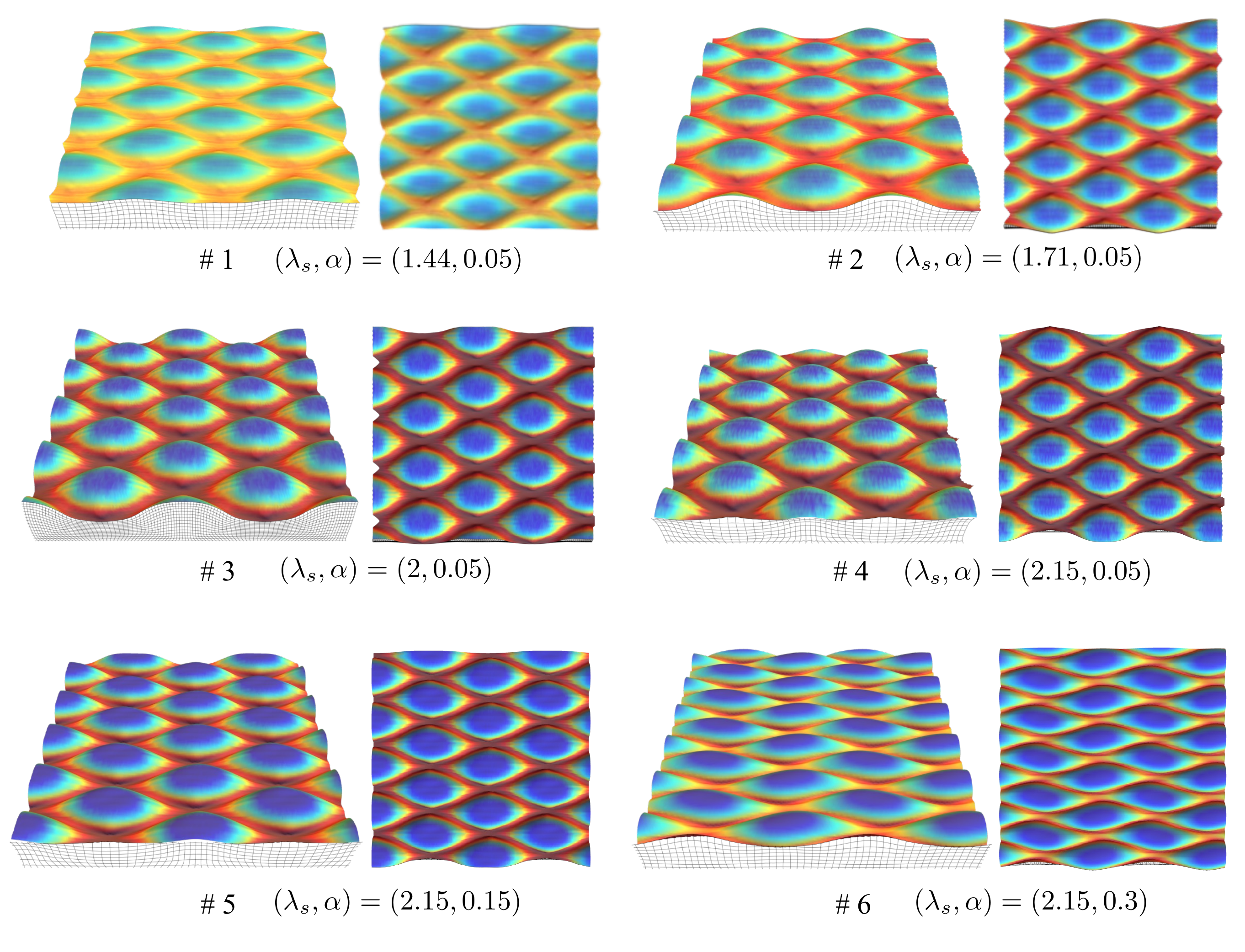}
	\caption{Side view and top view of simulations in Table.~\ref{tab:fullLCE}.}\label{fig:simulations}
\end{figure}

We provide the simulation details for Fig. 4 in Table \ref{tab:fullLCE} and more views in Fig.~\ref{fig:simulations}. Simulations $\#$1, $\#$2, $\#$3, and $\#$4 were conducted to explore the impact of $\alpha$, while simulations $\#$4, $\#$5, and $\#$6 were conducted to examine the effect of $\lambda_s$. Simulation $\#$0 was performed with a coarse mesh and large domain to investigate the dynamics and coarsening of the instability shown in Fig. 4(a).

\begin{table}[ht]
\centering
\resizebox{0.5\textwidth}{!}{\begin{tabular}{c|cccccccc}
\hline
$\#$ & $\lambda_s$  & $\alpha$ & $L_x$ & $L_y$   & $L_z$ & $n_x$ & $n_y$ & $n_z$ \\ \hline
0                &  2 & 0.05                  &  0.3  & 0.312   &  0.05 &  91  &  93  &    15                   \\
1                & 1.44  & 0.05                  &  0.073  & 0.157  &  0.05  &  43  &  91  &   31                    \\
2                & 1.71  & 0.05                  &  0.075  &  0.155 &  0.05 &   45 &  93  &   31                    \\
3                & 2  & 0.05                  &  0.075  &  0.156& 0.05   &  51  &  107   &  35                    \\
4                & 2.15 & 0.05                  &   0.07 &  0.15&  0.05  &  41  &  87  &    31                      \\
5                & 2.15 & 0.15                  &  0.066  &  0.172&  0.05  & 41   & 101   &  31                       \\
6               & 2.15 & 0.3                   &  0.055  &  0.162&  0.05   &  34  &  98  &    31                     \\ \hline
\end{tabular}}
\caption{Simulation details for Fig.~(4) in the main text.}
\label{tab:fullLCE}
\end{table}

\section{Linear stability analysis}
Many biological tissues and fiber-reinforced composites are examples of transversely isotropic materials. Unlike isotropic materials, the mechanical response of transversely isotropic materials remains the same under rotations about a preferred axis. Therefore, a general strain-energy function with four invariants is required to model the mechanical response of such materials if they are incompressible \cite{destrade2013least}. The use of a strain-energy function with four invariants also makes the linear stability analysis more challenging due to the reduction of symmetry.
Non-ideal LCE is an example of a transversely isotropic material, with $\bfn_0$ as the preferred axis. In this section, we present some general results of linear stability anaysis for transversely isotropic materials and a semi-theoretical linear stability analysis for our LCE instability problem using the numerical relaxed LCE energy $W_{R2}$.

Let the elastic energy for a transversely isotropic material be $W_N(\bflambda)$.
Following the free top surface condition, we variate the deformation gradient to its first order $\bflambda = \bflambda^0 + \epsilon \bflambda^1$, where $\bflambda^0 = \bfI + \gamma \hat{\bfz} \otimes \hat{\bfz}$ is the base state. Then we perturb the energy at its base state and get 
\beq
W_N(\bflambda) = W_N(\bflambda_0) + \left.\frac{\partial^2 W_N}{\partial \Lambda_{ij} \partial \Lambda_{kl}} \right |_{\bflambda^0} \Lambda^1_{ij}\Lambda^1_{kl}. \label{eq:energy_perturb}
\eeq
Taking the variation of (\ref{eq:energy_perturb}) yields the balance of force equation $\partial_i P_{ij}=0$, where
\beq
P_{ij} = \left.\frac{\partial^2 W_N}{\partial \Lambda_{ij} \partial \Lambda_{kl}} \right |_{\bflambda^0}\bflambda^1_{kl}.
\eeq
Owing to the uniaxial symmetry of $W_N(\bflambda)$, many of the second-order derivatives are zero. To compute this, we may write a general energy form $W_{N}(\bflambda)$ for a transversely isotropic material as \cite{destrade2013least}
\beq
W_{N} ( \bflambda) = g\left(\Tr(\bflambda \bflambda^\T), \Tr[(\bflambda \bflambda^\T)^{-1}], \bflambda \bfn_0\cdot\bflambda\bfn_0, \bflambda^\T\bflambda \bfn_0\cdot\bflambda^\T\bflambda\bfn_0, \mathrm{Det} \bflambda \right),
\eeq
where $\bfn_0 = \bfe_1$ is the preferred direction. Here we define $\bfe_1=\hat{\bfx},~ \bfe_2=\hat{\bfy},~ \bfe_1=\hat{\bfz}$.
Let $W_N^{\bfe_i \otimes \bfe_j + \bfe_k \otimes \bfe_l}$ denote the second-order derivative $\frac{\partial^2 W_N}{\partial \Lambda_{ij}\partial \Lambda_{kl}}$. For example,  $W_N^{((1,1,0),(0,0,0),(0,0,0))}$ denotes the second-order derivative $\frac{\partial^2 W_N}{\partial \Lambda_{11}\partial \Lambda_{12}}$. 
Taking the second-order derivative directly yields 15 non-zero second-order derivatives corresponding to
\begin{align}
&\bfe_i\otimes \bfe_j + \bfe_k\otimes\bfe_l = \nonumber\\
&\begin{cases}
((2, 0, 0), (0, 0, 0), (0, 0, 0)), \quad ((1, 0, 0), (0, 1, 0), (0, 0, 
0)), \quad ((1, 0, 0), (0, 0, 0), (0, 0, 1)), \\ ((0, 2, 0), (0, 0, 0), (0, 0, 
0)), \quad ((0, 1, 0), (1, 0, 0), (0, 0, 0)), \quad((0, 0, 2), (0, 0, 0), (0, 0, 
0)), \\((0, 0, 1), (0, 0, 0), (1, 0, 0)), \quad((0, 0, 0), (2, 0, 0), (0, 0, 
0)),\quad ((0, 0, 0), (0, 2, 0), (0, 0, 0)), \\((0, 0, 0), (0, 1, 0), (0, 0, 
1)), \quad((0, 0, 0), (0, 0, 2), (0, 0, 0)), \quad((0, 0, 0), (0, 0, 1), (0, 1, 
0)), \\
((0, 0, 0), (0, 0, 0), (2, 0, 0)), \quad((0, 0, 0), (0, 0, 0), (0, 2, 
0)),\quad ((0, 0, 0), (0, 0, 0), (0, 0, 2)).
\end{cases}
\end{align}
To solve the variational equation $\partial_i P_{ij}=0$, we substitute a test perturbed field that exponentially decays in $z$ as $\bflambda^1 = \bfv(x,y)\exp(k\lambda z)$. The $xy$ field $\bfv(x,y)$ has a wave vector $\bfk$ along $\hat{\bfx}'$ that makes an oblique angle $\theta$ to $\hat{\bfx}$, resulting in 
\beq
\bfv(x,y) = c_h \cos(k x') \hat{\bfx}' + c_g  \cos(k x')\hat{\bfy}' + \sin(k x') \hat{\bfz}.  
\eeq
Substituting $\bflambda^1$ into the variational equation yields three equations linear in $c_h$ and $c_g$ but quadratic in $\lambda$, which are

Equation EL(1):
\beqs
&&\lambda^2 \left(\sin\theta(c_g \cos\theta + c_h\sin\theta) W_N^{((0, 0, 0), (0, 0, 0), (0, 2, 0))} + 
\cos\theta(c_h \cos\theta - c_g \sin\theta) W_N^{((0, 0, 0), (0, 0, 0), (2, 0, 0))} 
\right)  \nonumber \\ &&+ \lambda \left( 
\sin^2\theta (W_N^{((0, 0, 0), (0, 0, 1), (0, 1, 0))}  + W_N^{((0, 0, 0), (0, 1, 0), (0, 0, 1))} )
+\cos^2\theta (W_N^{((0, 0, 1), (0, 0, 0), (1, 0, 0))}  + W_N^{((1, 0, 0), (0, 0, 0), (0, 0, 1))} )
\right)  \nonumber\\ &&+ (-c_g \sin ^3\theta  \cos \theta -c_h \sin ^4\theta )W_N^{((0, 0, 0), (0, 2, 0), (0, 0, 0))} + (c_g \sin ^3\theta  \cos \theta -c_h \sin ^2\theta  \cos ^2\theta )W_N^{((0, 0, 0), (2, 0, 0), (0, 0, 0))} \nonumber\\&&+ (-c_g \sin \theta  \cos ^3\theta +c_g \sin ^3\theta  \cos \theta -2 c_h \sin ^2\theta  \cos ^2\theta ) W_N^{((0, 1, 0), (1, 0, 0), (0, 0, 0))} \nonumber\\&&+ (-c_g \sin \theta  \cos ^3\theta -c_h \sin ^2\theta  \cos ^2\theta ) W_N^{((0, 2, 0), (0, 0, 0), (0, 0, 0))} \nonumber\\&&+ (-c_g \cos \theta^3 \sin\theta-2 c_h \cos\theta^2 \sin \theta^2+c_g \cos\theta \sin\theta^3) W_N^{((1,0,0),(0,1,0),(0,0,0))} \nonumber\\&& +\cos \theta^3 (-c_h \cos\theta+c_g \sin\theta) W_N^{((2,0,0),(0,0,0),(0,0,0))}=0, 
\eeqs

Equation EL(2):
\beqs
&& \lambda^2\left( 
-\cos\theta (\cos\theta c_g+c_h \sin\theta) W_N^{((0,0,0),(0,0,0),(0,2,0))}+\sin\theta (\cos\theta c_h-c_g \sin\theta) W_N^{((0,0,0),(0,0,0),(2,0,0))}
\right) \nonumber \\
&&+ \lambda \cos\theta \sin\theta \left(-W_N^{((0,0,0),(0,0,1),(0,1,0))}-W_N^{((0,0,0),(0,1,0),(0,0,1))}+W_N^{((0,0,1),(0,0,0),(1,0,0))}+W_N^{((1,0,0),(0,0,0),(0,0,1))}\right) \nonumber\\
&& + (c_g \cos\theta^2 \sin\theta^2+c_h \cos\theta \sin\theta^3) W_N^{((0,0,0),(0,2,0),(0,0,0))}+(-c_h \cos\theta \sin\theta^3+c_g \sin\theta^4) W_N^{((0,0,0),(2,0,0),(0,0,0))}\nonumber\\ &&+(c_h \cos\theta^3 \sin\theta-2 c_g \cos\theta^2 \sin\theta^2-c_h \cos\theta \sin\theta^3) W_N^{((0,1,0),(1,0,0),(0,0,0))}\nonumber\\ &&+(c_g \cos\theta^4+c_h \cos\theta^3 \sin\theta) W_N^{((0,2,0),(0,0,0),(0,0,0))} \nonumber\\ &&+(c_h \cos\theta^3 \sin\theta-2 c_g \cos\theta^2 \sin\theta^2-c_h \cos\theta \sin\theta^3) W_N^{((1,0,0),(0,1,0),(0,0,0))} \nonumber\\ &&+\cos\theta^2 \sin\theta (-c_h \cos\theta+c_g \sin\theta) W_N^{((2,0,0),(0,0,0),(0,0,0))}=0,
\eeqs

Equation EL(3):
\beqs
&&\lambda^2 W_N^{((0,0,0),(0,0,0),(0,0,2))} \nonumber\\ 
&&+\lambda \left(-\sin\theta (c_g \cos\theta+c_h \sin\theta) W_N^{((0,0,0),(0,0,1),(0,1,0))}-c_g \cos\theta \sin\theta W_N^{((0,0,0),(0,1,0),(0,0,1))}\right. \nonumber \\&&-c_h \sin\theta^2 W_N^{((0,0,0),(0,1,0),(0,0,1))}-c_h \cos\theta^2 W_N^{((0,0,1),(0,0,0),(1,0,0))} \nonumber \\ &&\left.+c_g \cos\theta \sin\theta W_N^{((0,0,1),(0,0,0),(1,0,0))}+\cos\theta (-c_h \cos\theta+c_g \sin\theta) W_N^{((1,0,0),(0,0,0),(0,0,1))}\right)\nonumber\\
&&-\sin\theta^2 W_N^{((0,0,0),(0,0,2),(0,0,0))}-\cos\theta^2 W_N^{((0,0,2),(0,0,0),(0,0,0))}=0.
\eeqs

Generically, all 15 second-order derivatives are non-zero. After solving for the coefficients $c_h$ and $c_g$, the equilibrium equation $\partial_i P_{ij}=0$ (i.e., equations EL(1), EL(2) and EL(3)) yields a cubic polynomial in $\lambda^2$. By selecting the decaying solutions in $-y$, we may obtain three positive roots  $\lambda=\lambda_1,\lambda_2,\lambda_3$, three corresponding pairs of coefficients $(c_{g1},c_{h1})$,  $(c_{g2},c_{h2})$, $(c_{g3},c_{h3})$, and three perturbed fields, $\bflambda^1_{(1)}$, $\bflambda^1_{(2)}$, and $\bflambda^1_{(3)}$. The total field can be expressed as the linear combination:
\beq
\bflambda^1=c_1 \bflambda^1_{(1)} + c_2 \bflambda^1_{(2)} + c_3 \bflambda^1_{(3)}.
\eeq
Then we substitute the perturbation mode $\bflambda^1$ back into the energy density equation (\ref{eq:energy_perturb}), and integrate over a period $\hat{x}'\in[0,2\pi/k]$ to obtain

\beqs
&&\int_0^{2\pi/k}\left.\frac{\partial^2 W_N}{\partial \Lambda_{ij} \partial \Lambda_{kl}} \right |_{\bflambda^0} \Lambda^1_{ij}\Lambda^1_{kl} \d x'= \nonumber \\
&&\pi f_z'^2 (W_N^{((0,0,0),(0,0,0),(0,0,2))}/(2 k)+\pi (\sin\theta f_x'+\cos\theta f_y')^2 W_N^{((0,0,0),(0,0,0),(0,2,0))}/(2 k) \nonumber\\&& +\pi (\cos\theta f_x'-\sin\theta f_y')^2 W_N^{((0,0,0),(0,0,0),(2,0,0))}/(2 k)+\pi f_z \sin\theta (\sin\theta f_x'+\cos\theta f_y') W_N^{((0,0,0),(0,0,1),(0,1,0))}\nonumber\\&&+1/2 k \pi f_z^2 \sin\theta^2 W_N^{((0,0,0),(0,0,2),(0,0,0))}-\pi \sin\theta (\cos\theta f_y+f_x \sin\theta) f_z' W_N^{((0,0,0),(0,1,0),(0,0,1))} \nonumber\\ &&+1/2 k \pi \sin\theta^2 (\cos\theta f_y+f_x \sin\theta)^2 W_N^{((0,0,0),(0,2,0),(0,0,0))}+1/2 k \pi \sin\theta^2 (\cos\theta f_x-f_y \sin\theta)^2 W_N^{((0,0,0),(2,0,0),(0,0,0))} \nonumber\\ &&+\pi \cos\theta f_z (\cos\theta f_x'-\sin\theta f_y') W_N^{((0,0,1),(0,0,1),(1,0,0))}+1/2 k \pi \cos\theta^2 f_z^2 W_N^{((0,0,2),(0,0,0),(0,0,0))} \nonumber\\ &&+k \pi \cos\theta \sin\theta (\cos\theta f_y+f_x \sin\theta) (\cos\theta f_x-f_y \sin\theta) W_N^{((0,1,0),(1,0,0),(0,0,0))}\nonumber\\ &&+1/2 k \pi \cos\theta^2 (\cos\theta f_y+f_x \sin\theta)^2 W_N^{((0,2,0),(0,0,0),(0,0,0))} \nonumber\\&&-\pi \cos\theta (\cos\theta f_x-f_y \sin\theta) f_z' W_N^{((1,0,0),(0,0,0),(0,0,1))} \nonumber\\ &&+k \pi \cos\theta \sin\theta (\cos\theta f_y+f_x \sin\theta) (\cos\theta f_x-f_y \sin\theta) W_N^{((1,0,0),(0,1,0),(0,0,0))} \nonumber\\ &&+1/2 k \pi \cos\theta^2 (\cos\theta f_x-f_y \sin\theta)^2 W_N^{((2,0,0),(0,0,0),(0,0,0))}
\eeqs
where $f_x(z) = c_1 c_{h1}\exp(k\lambda_1 z) +  c_2 c_{h2}\exp(k\lambda_2 z) +  c_3 c_{h3}\exp(k\lambda_3 z) $, $f_y(z) = c_1 c_{g1}\exp(k\lambda_1 z) +  c_2 c_{g2}\exp(k\lambda_2 z) +  c_3 c_{g3}\exp(k\lambda_3 z) $ and $f_z(z) = c_1\exp(k\lambda_1 z) +  c_2exp(k\lambda_2 z) +  c_3 \exp(k\lambda_3 z) $. Then integrating upon the depth $z \in (-\infty, 0]$ yields a quadratic form of the coefficients $c_1$, $c_2$, and $c_3$.
To determine stability, we calculate the minimum eigenvalue of the resulting Hessian matrix. A negative eigenvalue indicates instability. In the range $\theta \in [0, \pi/2]$, the smallest negative eigenvalue corresponds to the fastest decay in energy. 

{\bf LCE + neo-Hookean case}.  The combination of LCE and neo-Hookean energy has the form 
\beq
W_N(\bflambda) = f W_{R2}(\bflambda/\mathrm{Det}\bflambda) + (1-f)W_{NH}(\bflambda/\mathrm{Det}\bflambda) + B(\mathrm{Det} \bflambda - 1)^2/2.
\eeq
We consider an example with energy at the critical volume fraction $f=f_c=0.963$.
The base state $\bflambda_0=\bfI + \gamma \hat{\bfz} \otimes \hat{\bfz}$ satisfies the top free surface condition, i.e., 
$\left.\frac{\partial W_N}{\partial \bflambda}\right|_{\bflambda = \bflambda_0}\hat{\bfz}=0$, resulting in $\gamma = -0.0006544$. We then evaluate the derivatives at the base state:
\begin{align}
\begin{cases}
W_N^{((2, 0, 0), (0, 0, 0), (0, 0, 0))}=100.2039, \quad W_N^{((1, 0, 0), (0, 1, 0), (0, 0, 
0))}=99.7018, \quad W_N^{((1, 0, 0), (0, 0, 0), (0, 0, 1))}=99.7671, \\ W_N^{((0, 2, 0), (0, 0, 0), (0, 0, 
0))}=0.2333, \quad W_N^{((0, 1, 0), (1, 0, 0), (0, 0, 0))}=0.2332, \quad W_N^{((0, 0, 2), (0, 0, 0), (0, 0, 
0))}=0.2336, \\ W_N^{((0, 0, 1), (0, 0, 0), (1, 0, 0))}=0.2334, \quad W_N^{((0, 0, 0), (2, 0, 0), (0, 0, 
0))}=0.03702,\quad W_N^{((0, 0, 0), (0, 2, 0), (0, 0, 0))}=99.8917, \\W_N^{((0, 0, 0), (0, 1, 0), (0, 0, 
1))}=99.8831, \quad W_N^{((0, 0, 0), (0, 0, 2), (0, 0, 0))} = 0.03702, \quad W_N^{((0, 0, 0), (0, 0, 1), (0, 1, 
0))}=0.03699, \\
W_N^{((0, 0, 0), (0, 0, 0), (2, 0, 0))}=0.03702, \quad W_N^{((0, 0, 0), (0, 0, 0), (0, 2, 
0))}=0.03702,\quad W_N^{((0, 0, 0), (0, 0, 0), (0, 0, 2))}=100.0226.
\end{cases}
\end{align}

The perturbed field at $\theta=0.327\pi$ is
\beqs
\bflambda^1_{(1)}&=&1.0429 \exp(1.0430 z) \hat{\bfx}'-0.1377 \exp(1.0430 z) \hat{\bfy}'+ \exp(1.0430 z) \hat{\bfz}, \nonumber \\
\bflambda^1_{(2)}&=& 1.4182 \exp(1.4181 z) \hat{\bfx}' -1.1722 \exp(1.4181 z) \hat{\bfy}' + \exp(1.4181 z) \hat{\bfz}, \\ \nonumber
\bflambda^1_{(3)}&=& 1.5578 \exp(1.5568 z) \hat{\bfx}' +0.9471 \exp(1.5568 z) \hat{\bfy}' + \exp(1.5568 z) \hat{\bfz}.
\eeqs
The energy expression obtained is quadratic and can be written as follows:
\begin{equation}
E=0.5341 c_1^2+1.0990 c_2^2+c_1(1.5216c_2+0.9636 c_3)+1.2808 c_2 c_3+0.5717 c_3^2.
\end{equation}
To determine stability, we calculate the minimum eigenvalue of the Hessian matrix, which is found to be $-0.00009$. This value is just below zero, indicating a marginally unstable system.

{\bf Pure LCE case}. In the pure LCE case, owing to the evolution of the second-order twin, three additional second-order derivatives, $W_N^{((0, 0, 0), (0, 0, 2), (0, 0, 
0))}$, $W_N^{((0, 0, 0), (0, 0, 0), (0, 2, 
0))}$ and $W_N^{((0, 0, 0), (0, 0, 1), (0, 1, 
0))}$ are zero. The underlying mechanism is that by forming the second-order twin, the shear $(\bfI + s (\bfn_0 \times \bfn_0^{\perp}) \otimes\bfn_0^{\perp} )$ will induce changes of volume fractions but preserve the energy density.
This reduces the aforementioned cubic polynomial of $\lambda^2$ to a quadratic equation of $\lambda^2$. Then solving the equation yields the perturbed field $\bflambda^1$ being the combination of two decaying fields.

For example, in the $\lambda_s=2, \alpha=0.05$ case, the 12 non-zero second-order derivatives of the energy 
\beq
W_N(\bflambda) =  W_{R2}(\bflambda/\mathrm{Det}\bflambda)  + B(\mathrm{Det} \bflambda - 1)^2/2
\eeq
evaluated at the base state $\bflambda = \bfI -0.0006792 \hat{\bfz} \otimes \hat{\bfz}$
are
\begin{align}
\begin{cases}
W_N^{((2, 0, 0), (0, 0, 0), (0, 0, 0))}=100.17212, \quad W_N^{((1, 0, 0), (0, 1, 0), (0, 0, 
0))}=99.7121, \quad W_N^{((1, 0, 0), (0, 0, 0), (0, 0, 1))}=99.7793, \\ W_N^{((0, 2, 0), (0, 0, 0), (0, 0, 
0))}=0.2118, \quad W_N^{((0, 1, 0), (1, 0, 0), (0, 0, 0))}=0.2117, \quad W_N^{((0, 0, 2), (0, 0, 0), (0, 0, 
0))}=0.2121, \\ W_N^{((0, 0, 1), (0, 0, 0), (1, 0, 0))}=0.2119, \quad W_N^{((0, 0, 0), (2, 0, 0), (0, 0, 
0))}=0.0100,\quad W_N^{((0, 0, 0), (0, 2, 0), (0, 0, 0))}=99.8512, \\W_N^{((0, 0, 0), (0, 1, 0), (0, 0, 
1))}=99.8985, \quad
W_N^{((0, 0, 0), (0, 0, 0), (2, 0, 0))}=0.0100, \quad W_N^{((0, 0, 0), (0, 0, 0), (0, 0, 2))}=99.9858,
\end{cases}
\end{align}
resulting in two perturbed fields
\beqs
\bflambda^1_{(1)} &=& 0.77817555 \exp(0.77827228 z) -0.20666114 \exp(0.77827228 z) + \exp(0.77827228 z) ,\nonumber \\
\bflambda^1_{(2)} &=& 2.029273 \exp(2.0298365 z) -4.6718188 \exp(2.0298365 z)  + \exp(2.0298365 z)
\eeqs
and the minimum eigenvalue of Hessian being $-0.239$ at $\theta=0.35\pi$.

\section{Movie captions}
{\bf M1}. 85 $\mu$m thick film cooling from the isotropic state under reflection light.

{\bf M2}. 220 $\mu$m thick film cooling from the isotropic state under reflection light.

{\bf M3}. 220 $\mu$m thick film heating and cooling cycles under transmission POM.

{\bf M4}. 85 $\mu$m thick film cooling from the isotropic state under transmission POM. Cross-hatch microstructure evolution is seen upon cooling.

{\bf M5}. Simulation of the dynamics and coarsening of surface instability in a nematic elastomer.

Supporting videos are available at \url{https://drive.google.com/drive/folders/18DCJz0DnNKFKmrJ1Ggy\_Q40\_UhXbj8XH?usp=sharing}

\bibliography{elastomer}